\documentclass[a4paper,fleqn]{cas-dc}

\usepackage{amsmath,amsfonts}
\usepackage{algorithm}
\usepackage{algorithmic}
\usepackage{tikz}
\usetikzlibrary{arrows.meta,positioning,fit,shadows.blur,shapes.multipart}
\usepackage{booktabs}
\usepackage{multirow}
\usepackage{threeparttable}
\usepackage{float}

\usepackage{stfloats}
\usepackage{placeins}

\usepackage{rotating}
\usepackage{threeparttable}
\usepackage{array}
\usepackage[caption=false,font=normalsize,labelfont=sf,textfont=sf]{subfig}

\usepackage{xcolor}
\usepackage{cite}
\usepackage{hyperref}

\usepackage[numbers]{natbib}
\usepackage[normalem]{ulem}
\useunder{\uline}{\ul}{}
\usepackage{colortbl}
\usepackage[utf8]{inputenc}
\usepackage{tcolorbox}
\usepackage{tabularx}

\def\tsc#1{\csdef{#1}{\textsc{\lowercase{#1}}\xspace}}
\tsc{WGM}
\tsc{QE}
\tsc{EP}
\tsc{PMS}
\tsc{BEC}
\tsc{DE}

\begin{document}

\ExplSyntaxOn
\cs_set:Npn \__first_footerline:
{
  \group_begin:
  \small\sffamily
  \ifnum\theblind>0\relax\else \__short_authors: \fi
  \group_end:
}
\ExplSyntaxOff

\shorttitle{}
\shortauthors{Abhishek Kumar et~al.}

\title [mode = title]{ LinkRank: A Learning-to-Rank Framework for One-to-Many Issue–Commit Traceability}                      

\author[1]{Abhishek Kumar}
\ead{akumar@turing.ac.uk}

\author[2]{Tuhin Mondal}
\ead{email2tuhin04@gmail.com}

\author[3]{Partha Pratim Das}
\ead{ppd@ashoka.edu.in}

\author[2]{Partha Pratim Chakrabarti}
\ead{ppchak@cse.iitkgp.ac.in}

\affiliation[1]{organization={The Alan Turing Institute},
                city={London},
                country={United Kingdom}}
                
\affiliation[2]{organization={Department of Computer Science \& Engineering, Indian Institute of Technology},
                city={Kharagpur},
                country={India}}

\affiliation[3]{organization={Department of Computer Science, Ashoka University},
                city={Sonipat},
                country={India}}

\begin{abstract}
Recovering traceability links between issues and commits is important for software maintenance, debugging, impact analysis, and project understanding. However, most existing approaches assume a one-to-one relationship, where each issue is linked to a single commit. In practice, many issues are resolved through multiple commits, and ignoring this one-to-many nature can lead to incomplete traceability. This paper presents \textit{LinkRank}, a learning-to-rank framework for recovering one-to-many issue--commit links. Unlike existing methods that mainly judge issue--commit pairs independently, LinkRank considers the set of candidate commits for an issue and identifies the commits that are most likely to contribute to its resolution. To support realistic evaluation, we construct a new dataset from six open-source GitHub repositories. LinkRank follows an iterative pick--remove--renormalize strategy: it selects the highest-ranked commit, removes it from the candidate pool, renormalizes the remaining scores, and repeats the process until the stopping criterion is met. We evaluate LinkRank under two settings: Known-$K$, where the true number of linked commits is available, and Unknown-$K$, where the model must infer when to stop selecting commits using ABS and REL stopping rules. Across six projects, LinkRank achieves an average Known-$K$ F1 score of 74.54\%, compared with 48.39\% for the strongest baseline. In the Unknown-$K$ setting, LinkRank achieves 68.84\% F1 with ABS and 67.02\% F1 with REL, outperforming the strongest baselines under both automatic stopping rules. Overall, the findings suggest that one-to-many issue--commit traceability is better addressed as an issue-centric ranking and iterative selection problem than as independent pairwise classification.
\end{abstract}


\maketitle

\section{Introduction}

Software traceability, the process of establishing and 
maintaining links between artifacts such as requirements, 
issues, commits, test cases, and design documents 
\cite{traceability, intro_testing}, underpins a wide 
range of software engineering activities, from change 
impact analysis \cite{impact_analysis} and bug fixing to 
safety-critical development \cite{safety}, project 
management \cite{project}, and long-term software 
maintenance. By making explicit how artifacts relate and 
how a change in one propagates to others, these links give 
developers the context they need to reason about evolving 
systems. Among different forms of traceability, issue--commit linking is particularly important because it connects reported issues with the code changes made to resolve them \cite{hermes, ealink, pi_issue_pr}, making it possible to understand not just \textit{what} 
changed in a codebase, but \textit{why}.

In modern software projects, issues are usually managed in issue tracking systems such as \textit{Bugzilla} \cite{r1} and \textit{Backlog} \cite{r2}, while commits are maintained in version control systems such as \textit{Git} \cite{r3} and \textit{Mercurial} \cite{r4}. Although developers can manually refer to issue identifiers in commit messages, this practice is often inconsistent or incomplete. As a result, many issue--commit links remain missing, which makes it harder to understand why a change was introduced, which issue it addressed, and how the project evolved over time \cite{re1}. Missing links can increase maintenance effort and affect downstream software engineering tasks such as bug prediction \cite{r6}, commit analysis \cite{r5}, debugging, and impact analysis \cite{cost1, brindescu2014centralized, cost3}.

To recover such missing links, researchers have proposed a wide range of automated approaches. Early methods relied mainly on heuristics, textual similarity, and manually designed rules \cite{r6, r48, rene1}. Later approaches introduced machine learning models that used textual and metadata-based features to classify issue--commit pairs \cite{q2, rene2, q4, q3, r56}. More recently, deep learning and large language model (LLM) based models have been used to capture richer semantic relationships between issues and commits \cite{rene3, q1, ealink, btlink, traceability, mplinker, akhavan2025linkanchor, huang2025back}. These approaches have improved the state of issue--commit link recovery and have shown the value of using learned representations for traceability tasks.

\begin{figure}
  \centering
  \includegraphics[width=0.99\linewidth]{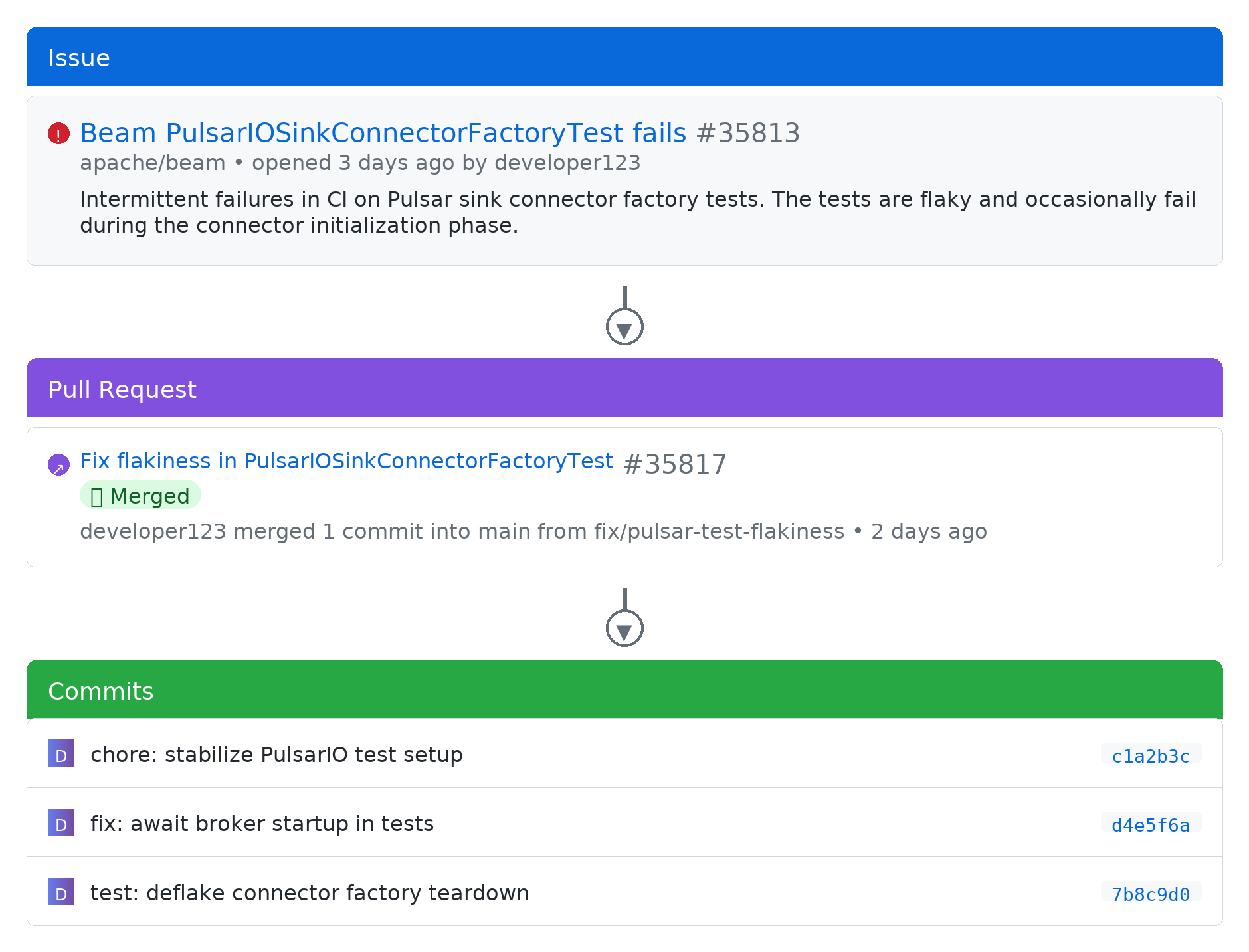}
  \caption{An example of one-to-many issue--commit traceability, where a single issue (\#35813) is resolved through a pull request (\#35817) containing multiple focused commits. Each commit contributes to a different part of the issue resolution, such as test setup stabilization, broker startup synchronization, and teardown improvement.}
  \label{intro}
\end{figure}

However, most existing studies formulate issue--commit link recovery as a one-to-one problem, where each issue is assumed to be linked to a single commit. This assumption does not fully reflect real software development practice. In practice, a single issue may be resolved through several commits, especially when the change requires multiple implementation steps, test updates, refactoring, or follow-up fixes. For example, Figure~\ref{intro} shows a GitHub issue that is resolved through a pull request containing multiple related commits. Each commit contributes to a different part of the resolution. Treating such cases as independent one-to-one links can lead to incomplete traceability because the model may recover only one relevant commit while missing other commits that are also necessary to understand the full resolution of the issue.

To address these limitations, we propose \textit{LinkRank}, 
a learning-to-rank framework for recovering one-to-many 
issue--commit links. Unlike existing methods that score 
each issue--commit pair in isolation, LinkRank is built on 
a different premise: one-to-many traceability is a 
set-recovery problem, not a classification problem. The 
goal is therefore not to decide whether a single commit is 
linked to an issue, but to recover the complete set of 
commits that jointly explain its resolution. To achieve 
this, LinkRank treats each issue as a query over a pool of 
candidate commits and applies an iterative selection 
process: it selects the most relevant commit, removes it 
from the candidate pool, renormalizes the remaining scores, 
and repeats until the stopping criterion is met. This 
formulation is well suited to the one-to-many setting 
because it allows the model to compare candidates within 
the same issue-specific pool and decide which commits 
together form the resolution set, something independent 
pairwise classifiers cannot do by design.

We evaluate LinkRank on a newly constructed dataset from 
six open-source GitHub repositories with realistic candidate pools 
in which relevant commits are sparse among hundreds of 
candidates. Our results show that LinkRank consistently 
outperforms recent issue--commit linking baselines across 
all evaluated projects, achieving a Known-$K$ F1 of 
73.17\% against 48.39\% for the strongest baseline, a 
gap that holds even against recent LLM-based approaches 
with far greater model capacity. In the more realistic 
Unknown-$K$ setting, where the number of linked commits 
must be inferred automatically, LinkRank reaches 65.94\% 
F1 with ABS stopping and 62.19\% with REL, outperforming 
all baselines under both rules. Together, these results 
suggest that correctly framing the task as set recovery 
matters more than model complexity: a well-formulated 
lightweight ranker consistently outperforms heavier neural 
and LLM-based methods designed for one-to-one link 
recovery.

In summary, this paper makes the following contributions:
\begin{enumerate}

\item \textit{A problem reframing of one-to-many 
issue--commit traceability.} We show that one-to-many 
issue--commit recovery is fundamentally an issue-level 
set-recovery problem rather than independent pairwise 
binary classification. This reframing is central to 
LinkRank, which compares candidate commits within the 
same issue-specific pool and iteratively selects the 
commits that jointly explain the issue resolution.

\item \textit{A realistic dataset and evaluation setting 
for one-to-many issue--commit traceability.} We construct 
a new dataset from six open-source GitHub repositories 
that preserves the complete set of commits associated with 
each issue, covering 3,103 issues and 7,688 commits across 
diverse domains. The dataset includes both one-to-one and 
one-to-many cases and is evaluated using realistic 
candidate pools where relevant commits are sparse among 
hundreds of unrelated candidates.

\item \textit{Empirical evidence that problem formulation 
matters more than model complexity.} We compare LinkRank 
against four baselines, including recent neural and 
LLM-based approaches demonstrating that a well-formulated lightweight 
ranker consistently outperforms heavier methods designed 
for one-to-one link recovery.

\end{enumerate}

The remainder of this paper is organized as follows. Section~2 reviews related work on issue--commit link recovery. Section~3 presents the proposed LinkRank framework, including dataset construction, feature representation, LambdaMART-based ranking, and iterative selection strategies. Section~4 describes the experimental setup, evaluation metrics, and baseline methods. Section~5 reports the empirical results and analyzes LinkRank's effectiveness, ranking quality, feature contributions, and computational cost. Section~6 discusses threshold sensitivity, threats to validity, and future work. Finally, Section~7 concludes the paper.

\section{Related Work}

\textbf{Rule-based approaches.} Bird et al. \cite{r23} observed that developers often struggle with manually identifying missing links between issues and commits, a process that is both time-consuming and tedious. To assist developers in this task, they introduced LINKSTER, a system that provides query interfaces for both issue and commit data, making it easier to retrieve relevant links. However, LINKSTER still required manual effort, as it did not fully automate the link recovery process. To overcome this limitation, Wu et al. developed ReLink \cite{r6}, the first automatic approach for recovering missing links between issues and commits. ReLink leveraged textual similarity by treating issue reports and commit log messages as plain text, allowing for automated text feature extraction and comparison. While this marked a significant step forward, the reliance solely on textual similarity proved insufficient, as it failed to capture cases where issue descriptions and commit messages were semantically related but lexically different. Recognizing this shortcoming, Nguyen et al. \cite{r48} introduced MLINK, an approach that extended ReLink by incorporating both textual features and structural information from the modified source code. Unlike ReLink, which focused exclusively on textual similarity, MLINK considered actual code changes associated with bug fixes, enabling a more comprehensive and accurate recovery of missing links. By integrating both natural language processing techniques and software-specific modifications, MLINK significantly improved the precision of issue-commit link recovery.

\textbf{Machine-learning approaches.} Previous approaches, such as ReLink and MLink, relied on textual similarities between commit logs and issue reports, often failing when commit messages were vague or missing. RCLinker \cite{q3} addressed this by integrating ChangeScribe \cite{r27, r28}, which generates commit messages summarizing code changes. By combining these generated summaries with developers’ original messages, RCLinker enriched textual features and adopted a classification-based approach using a random forest classifier to predict issue-commit links. While RCLinker improved upon prior methods, it still overlooked non-source documents present in commits. FRLink \cite{r56} tackled this limitation by incorporating such documents while filtering out irrelevant files, enhancing contextual information beyond conventional text-based approaches. However, most existing models framed the problem as binary classification, disregarding unlabelled links. Recognizing this, PULink \cite{q4} reframed link recovery as a positive-unlabelled (PU) learning task, distinguishing unlabelled links from negative ones, leading to better model training and improved performance. Despite these advancements, prior models still suffered from low precision and high computational costs, particularly due to the scarcity of textual data. HybridLinker \cite{q2} addressed this by combining textual and non-textual information in a hybrid model. It trained separate classifiers for both data types and fused their outputs using a weighted ensemble technique, resulting in higher precision, lower computational cost, and improved link recovery accuracy.

\textbf{Deep-learning approaches.}
DeepLink~\cite{rene3,q1} is among the early deep-learning approaches for issue--commit link recovery. It uses a code knowledge graph to capture semantic information from code changes, moving beyond earlier methods that relied mainly on textual and code similarity. More recent transformer-based approaches further improved issue--commit link recovery by learning richer semantic representations. T-BERT~\cite{traceability} uses transfer learning to address data scarcity and computational cost, BTLink~\cite{btlink} employs dual BERT encoders to represent issue and commit information, and EALink~\cite{ealink} uses knowledge distillation and contrastive learning to improve efficiency and link prediction performance. However, these methods still mainly formulate issue--commit recovery as pairwise binary classification. Even when their datasets contain issues linked to multiple commits, they do not explicitly model the complete one-to-many commit set associated with an issue. Wang et al.~\cite{mplinker} proposed \textit{MPLinker}, a prompt-tuning-based approach for issue--commit link recovery. MPLinker reformulates link prediction as a cloze-style task, where prompt templates are used to predict whether an issue--commit pair is linked. It also combines multiple prompt templates with adversarial training to improve generalization.

\textbf{LLM-based approaches.} Huang et al.~\cite{huang2025back} proposed \textit{EasyLink}, an LLM-assisted retrieval approach for issue--commit linking. EasyLink constructs realistic candidate pools and uses dense retrieval followed by GPT-4o reranking. Although it improves retrieval under realistic settings, it mainly focuses on ranking or retrieving candidate commits using top-$k$ metrics. Akhavan et al.~\cite{akhavan2025linkanchor} proposed \textit{LinkAnchor}, an autonomous LLM-based agent for issue-to-commit link recovery. LinkAnchor uses tool calls to inspect repository context, including commit history, issue discussions, diffs, and code files, and reasons over related commits to identify the final resolving commit. Although it is multi-commit-aware during reasoning, its primary output is still a single final commit. 

Across all these approaches, rule-based, ML, deep learning, and LLM-based, the shared assumption is that issue–commit recovery is a pairwise decision problem. We argue this assumption is the root cause of poor multi-commit recovery, not the choice of model or features. The correct formulation treats each issue as a query over a ranked pool, and resolution as recovering a complete commit set.

\section{The LinkRank Approach}

LinkRank is designed around the set-recovery view of one-to-many issue--commit traceability. Since the goal is to recover a commit set for each issue, the framework must first build issue-level candidate pools, then rank candidates within each pool, and finally select the commits that jointly explain the issue resolution.

\begin{figure*}
  \centering
  \includegraphics[width=0.99\linewidth]{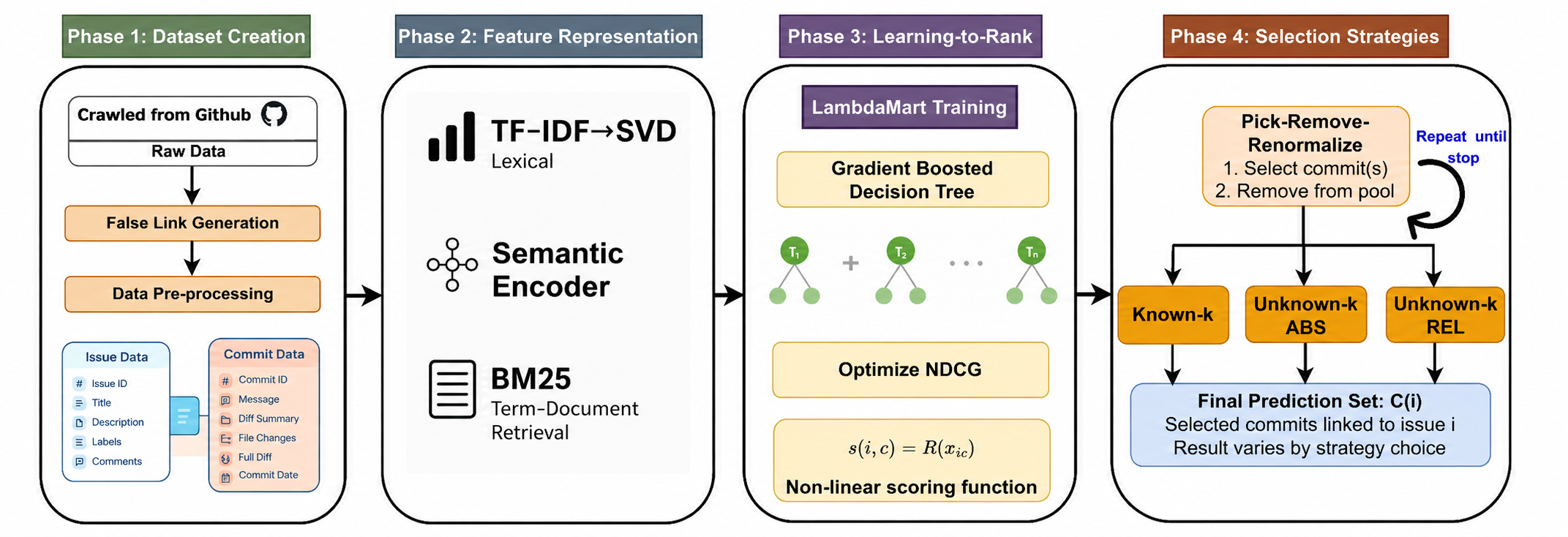}
  \caption{Overview of the proposed LinkRank framework for one-to-many issue--commit recovery.}

  \label{fig:approach}
\end{figure*}

Our approach is organized into four phases: (1) dataset construction, (2) feature representation, (3) learning-to-rank with LambdaMART, and (4) selection strategies. The complete workflow is illustrated in Figure~\ref{fig:approach}, and each phase is discussed in the following subsections.

\subsection{Phase I: Dataset Construction Process}

\subsubsection{Why a New Dataset Is Needed}
Existing issue--commit linking datasets are primarily 
designed for pairwise link recovery, where the task is 
to predict whether a given issue--commit pair is linked 
or not. As a result, data is typically organised as 
independent linked and unlinked pairs rather than as an 
issue-level structure where each issue is associated with 
its complete set of resolving commits, making such datasets inadequate for evaluating 
one-to-many traceability recovery.

Some recent datasets do contain issues linked to multiple 
commits. EALink~\cite{ealink}, for instance, includes 
pair-level examples where some issues correspond to 
multiple commits, but its released form is preprocessed 
and tokenized rather than provided as raw artifacts. 
Consequently, information essential for realistic 
one-to-many recovery, complete issue text, commit 
messages, diffs, timestamps, file paths, and the full 
per-issue commit set, is not directly accessible.

Our study therefore requires a dataset that preserves 
the issue-level structure of traceability links, with 
each issue associated with its complete ground-truth 
commit set and evaluated against a realistic candidate 
pool of competing commits from the same repository. We 
construct such a dataset from scratch, preserving raw 
issue and commit artifacts, ground-truth commit 
multiplicity, and realistic candidate pools that reflect 
the difficulty of one-to-many recovery in practice.

\subsubsection{Data Collection}

The dataset was constructed using GitHub's REST API to extract
issue--commit pairs from open-source repositories. GitHub imposes a
rate limit of 5{,}000 requests per hour per authenticated
user~\cite{GitHubAPI}, making large-scale extraction time-consuming.
We targeted repositories that (i)~are actively maintained with rich
issue-tracking practices, (ii)~contain a sufficient number of issues
linked to multiple commits, and (iii)~span diverse application domains
(data processing, RPC frameworks, data lakes, machine learning
frameworks) to improve generalizability.

\subsubsection{Inclusion Criteria}

To construct a dataset suitable for one-to-many issue--commit traceability recovery, we include issues linked to $K=1$ through $K=7$ commits. Including $K=1$ issues is important because one-to-one links remain common in practice and should not be excluded from a realistic benchmark. As shown in Table~\ref{stats}, $K=1$ issues constitute a substantial portion of each project, ranging from 26\% to 55\% of the dataset. Their inclusion allows the dataset to represent both traditional one-to-one cases and more complex one-to-many cases.

We set the upper bound to $K=7$ based on the observed distribution of linked commits in the selected repositories. As shown in Table~\ref{stats}, the average number of commits per issue ranges from approximately 2 to 4 across the repositories, indicating that most issues are resolved through a relatively small number of commits. Issues linked to a much larger number of commits are comparatively rare and often correspond to broad development activities, such as large-scale refactoring, automated changes, or pull requests that combine several loosely related modifications. Including such cases may introduce noise and make it difficult to determine whether all commits are directly relevant to the issue resolution. Therefore, the range $K=1$ to $K=7$ provides a balance between realism and dataset quality: it captures common multi-commit issue resolutions while avoiding very high-$K$ cases that may not represent focused traceability links.

We initially surveyed a set of candidate repositories and selected six projects based on three criteria: the availability of issue--commit links, the density of one-to-many cases, and sufficient scale for cross-validation. The selected projects also cover different software domains, including data processing, query engines, RPC frameworks, table formats, and machine learning frameworks. The surveyed repositories are shown in Table~\ref{stats_survey}, and the final dataset statistics are summarized in Table~\ref{stats}.

\begin{table*}[htbp]
\centering
\caption{Survey of candidate repositories. The six projects were selected for evaluation based on the availability of issue--commit links, the density of one-to-many cases, and domain diversity.}
\renewcommand{\arraystretch}{1.3}
\setlength{\tabcolsep}{6pt}
\label{stats_survey}
\scriptsize
\begin{tabular}{lrcrc}
\hline
\textbf{Project} & \textbf{Stars} & \textbf{Domain} &
\textbf{Issues (2$\leq$K$\leq$7)} & \textbf{Selected?} \\ \hline
\textbf{Apache Beam}       & 8.3k  & Data processing   & 390 & \checkmark \\
\textbf{Apache DataFusion} & 7.7k  & Query engine       & 543 & \checkmark \\
\textbf{Apache Dubbo}      & 41.3k & RPC framework      & 209 & \checkmark \\
\textbf{Apache Iceberg}    & 7.9k  & Table format       & 305 & \checkmark \\
\textbf{Apache MXNet}      & 20.7k & ML framework       & 204 & \checkmark \\
\textbf{PyTorch}           & 93.2k & ML framework       & 135 & \checkmark \\\hline
\end{tabular}
\end{table*}

\begin{table*}[htbp]
\centering
\caption{Statistics of the final dataset (K\,=\,1..7). Each issue is
linked to exactly $K$ commits. The ``Candidate Pool'' column shows the
average number of candidate commits per issue in the RDS window,
reflecting the realistic retrieval difficulty.}
\renewcommand{\arraystretch}{1.4}
\setlength{\tabcolsep}{5pt}
\label{stats}
\scriptsize
\begin{tabular}{lrrrrrrrrrrrr}
\hline
\multirow{2}{*}{\textbf{Project}} &
\multirow{2}{*}{\textbf{Issues}} &
\multirow{2}{*}{\textbf{Commits}} &
\multirow{2}{*}{\textbf{Candidate Pool}} &
\multicolumn{7}{c}{\textbf{Number of Issues with K Commits}} &
\multirow{2}{*}{\textbf{Avg K}} \\ \cline{5-11}
 & & & & \textbf{1} & \textbf{2} & \textbf{3} & \textbf{4} &
   \textbf{5} & \textbf{6} & \textbf{7} & \\ \hline
Apache Beam       & 671 & 1{,}625 & $\sim$590     & 281 & 144 &  92 &  62 & 42 & 28 & 22 & 2.42 \\
Apache DataFusion & 738 & 2{,}270 & $\sim$1{,}264 & 195 & 130 & 139 & 106 & 73 & 56 & 39 & 3.08 \\
Apache Dubbo      & 469 &   938 & $\sim$133     & 260 &  93 &  50 &  21 & 20 & 16 &  9 & 2.00 \\
Apache Iceberg    & 551 & 1{,}357 & $\sim$193     & 246 & 107 &  62 &  41 & 43 & 31 & 21 & 2.46 \\
Apache MXNet      & 383 &   903 & $\sim$427     & 179 &  70 &  49 &  36 & 19 & 12 & 18 & 2.36 \\
PyTorch           & 291 &   595 & $\sim$439     & 156 &  52 &  45 &  14 &  9 &  6 &  9 & 2.04 \\ \hline
\textbf{Total}    & \textbf{3{,}103} & \textbf{7{,}688} & &
  \textbf{1{,}317} & \textbf{596} & \textbf{437} & \textbf{280} &
  \textbf{206} & \textbf{149} & \textbf{118} & \textbf{2.48} \\ \hline
\end{tabular}
\end{table*}

\subsubsection{Candidate Pool Construction}

A key requirement for realistic issue--commit link recovery is the construction of candidate pools that reflect the difficulty of the task in real repositories. Prior work often constructs negative examples by randomly sampling unrelated commits~\cite{btlink, q1}, using time-based pairing~\cite{hermes, ealink}, or generating large sets of unverified negative pairs~\cite{traceability}. While these strategies are useful for training and evaluation, they may simplify the task in two ways. First, randomly sampled negative commits can be too easy to distinguish from true links, which may inflate model performance. Second, balanced positive-to-negative ratios do not reflect real development settings, where only a small number of commits are relevant to a given issue among many possible candidates.

To address this, we construct candidate pools using a \emph{Relative Date Span} (RDS) strategy. For each issue~$i$ with ground-truth commit set $C_i = \{c_1, \dots, c_K\}$, we first identify the earliest and latest timestamps among its linked commits. We then include all commits from the same repository that fall within this time span, extended by a margin $\delta$ on both sides. Formally, the candidate pool for issue~$i$ is defined as:
\[
\mathcal{P}_i = \bigl\{ c \in \mathcal{C}_{\text{repo}} \;\big|\;
  t_c \in [t_{\min}(C_i) - \delta,\; t_{\max}(C_i) + \delta] \bigr\},
\]
where $t_{\min}(C_i)$ and $t_{\max}(C_i)$ denote the earliest and latest timestamps among the ground-truth commits for issue~$i$. In our experiments, we set the temporal margin $\delta$ to 365 days, creating a broad repository-specific candidate pool around the issue resolution period. Each candidate commit $c \in \mathcal{P}_i$ is then assigned a binary relevance label:
\[
y_{i,c} =
\begin{cases}
1 & \text{if } c \in C_i, \\
0 & \text{otherwise}.
\end{cases}
\]

This construction creates a realistic retrieval setting because the candidate pool contains commits that are temporally close to the actual resolution period of the issue. Such commits are more challenging negatives than randomly selected commits, since they may come from the same development period and may be related to similar project activities. As shown in Table~\ref{stats}, the resulting candidate pools contain, on average, 133 to 1{,}264 commits per issue. This leads to highly imbalanced positive-to-negative ratios, approximately between 1:65 and 1:410, reflecting the practical difficulty of identifying a small number of relevant commits from a large set of candidates.

The full evaluation dataset for a project is therefore defined as:
\[
\mathcal{D} = \bigcup_{i \in \mathcal{I}} \bigl\{
  (i, c, y_{i,c}) \;\big|\; c \in \mathcal{P}_i
\bigr\}.
\]

For model training, we use the complete candidate pool for each issue, including all positive and negative candidates within the RDS window. This allows the ranking model to learn the relative ordering of commits within each issue-specific candidate pool, rather than learning from isolated issue--commit pairs.

\subsection{Phase II: Feature Representation}

For each issue--commit candidate pair $(i,c)$, LinkRank 
constructs a feature vector $\mathbf{x}_{i,c} \in 
\mathbb{R}^d$ that captures different indicators of 
relevance between commit $c$ and issue $i$. We use four groups of features: lexical similarity, retrieval-based matching, semantic similarity, and metadata-based context. Together, these groups produce 17 features. The purpose of this multi-view representation is to combine complementary signals, since true issue--commit links may be indicated by textual overlap, semantic relatedness, temporal proximity, file-level overlap, or a combination of these factors.

\subsubsection{Lexical Similarity}

Lexical similarity captures surface-level textual overlap between issues and commits. We represent issue and commit texts using TF-IDF and then apply truncated Singular Value Decomposition (SVD) to obtain compact dense representations. The similarity between an issue and a commit is computed using cosine similarity between their SVD vectors:
\[
f_{\text{svd}}(i,c) = \frac{\langle \mathbf{v}_i, \mathbf{v}_c \rangle}{\|\mathbf{v}_i\|\cdot \|\mathbf{v}_c\|}.
\]
We compute this similarity over four textual views: full text, code-only content, natural language text, and title-level text. These views allow the model to capture different kinds of lexical evidence, such as overlap between issue descriptions and commit messages or between issue content and changed code.

\subsubsection{Retrieval-Based Matching}

In addition to TF-IDF-based similarity, we use BM25~\cite{bm25} as a retrieval-oriented matching signal. BM25 is useful because it gives higher weight to discriminative terms while accounting for document length and repeated term occurrences. In our setting, the issue text is treated as a query and the commit text as a document. As with the lexical features, BM25 scores are computed over four textual views: full text, code-only content, natural language text, and title-level text. These features capture direct term-matching evidence that may not be fully represented by dense SVD-based similarity.

\subsubsection{Semantic Similarity}

Lexical matching alone may miss relevant links when the issue and commit describe the same change using different words. To capture semantic relatedness, we use Sentence-BERT (SBERT)~\cite{reimers2019sentence} with the \texttt{all-mpnet-base-v2} model. For each issue and commit, we compute sentence embeddings and measure their cosine similarity:
\[
f_{\text{sbert}}(i,c) = \cos\bigl(\phi(i), \phi(c)\bigr).
\]
We compute SBERT similarity using three views: full text, natural language text, and title-level text. These features help identify links where the issue and commit are semantically related even when they have limited exact word overlap.

\subsubsection{Metadata Features}

In addition to textual and semantic similarity, LinkRank uses six metadata features to capture temporal and structural context around each issue--commit pair. These features are useful because relevant commits are not identified only by textual similarity; they may also be indicated by when the commit was made, which files it changed, and how large the change was.

\begin{itemize}
  \item \textbf{Temporal proximity} ($f_{\text{time}}$): The absolute time difference, in seconds, between the issue creation date and the commit author date. Commits related to an issue are often made close to the issue discussion or resolution period.

  \item \textbf{Temporal direction} ($f_{\text{dir}}$): The signed time difference between the issue creation date and the commit author date. This feature captures whether the commit was made before or after the issue was created.

  \item \textbf{Commit message length} ($f_{\text{desc}}$): The number of characters in the commit message. This can provide a weak signal about how much explanatory information the commit contains.

  \item \textbf{Diff size} ($f_{\text{diff}}$): The number of characters in the commit diff. This captures the scale of the code change, which may help distinguish focused fixes from very small or very large changes.

  \item \textbf{Number of files changed} ($f_{\text{nfiles}}$): The number of files modified by the commit. This provides information about the scope of the change.

  \item \textbf{File overlap} ($f_{\text{file}}$): The Jaccard similarity between the set of filenames mentioned in the issue text and the set of files modified in the commit:
    \[
    f_{\text{file}}(i,c) = \frac{|F_i \cap F_c|}{|F_i \cup F_c|},
    \]
    where $F_i$ is extracted from the issue text using path-like pattern matching and $F_c$ is obtained from the commit diff headers. This feature is especially useful when an issue explicitly refers to files, modules, or paths that are later modified by the commit.
\end{itemize}

These metadata features provide information that is not directly captured by text-based similarity measures. As shown in Section~\ref{results}, removing metadata features leads to the largest performance drop, indicating that temporal and file-level context plays an important role in one-to-many issue--commit recovery.

\subsubsection{Feature Summary}

Table~\ref{tab:features} summarizes the feature groups used in LinkRank. The multi-view design allows the model to consider complementary evidence from different sources. Lexical features capture surface-level term overlap, BM25 provides retrieval-based matching scores, SBERT captures semantic relatedness, and metadata features encode temporal and structural context that may not be visible from text alone. Together, these features provide a richer representation of each issue--commit candidate pair.

\begin{table}[htbp]
\centering
\caption{Feature groups used in LinkRank.}
\renewcommand{\arraystretch}{1.15}
\setlength{\tabcolsep}{4pt}
\label{tab:features}
\scriptsize
\begin{tabularx}{\columnwidth}{lXc}
\hline
\textbf{Group} & \textbf{Features} & \textbf{Count} \\ \hline
SVD & svd\_full, svd\_code, svd\_nl, svd\_title & 4 \\
BM25 & bm25\_full, bm25\_code, bm25\_nl, bm25\_title & 4 \\
SBERT & sbert\_full, sbert\_nl, sbert\_title & 3 \\
Metadata & time, time\_dir, desc\_len, diff\_size, nfiles, file\_jaccard & 6 \\ \hline
\textbf{Total} & & \textbf{17} \\ \hline
\end{tabularx}
\end{table}

\subsection{Phase III: Learning-to-Rank}\label{sec:ranking-model}

We formulate one-to-many issue--commit recovery as a learning-to-rank problem. For each issue $i$, the model receives a candidate pool $\mathcal{P}_i$ and assigns a relevance score to each candidate commit $c \in \mathcal{P}_i$. The goal is to rank commits that truly contribute to the resolution of the issue above unrelated commits. Formally, LinkRank learns a scoring function
\[
f(\mathbf{x}_{i,c}) \rightarrow \mathbb{R},
\]
where $\mathbf{x}_{i,c}$ is the feature vector representing the issue--commit pair $(i,c)$. A higher value of $f(\mathbf{x}_{i,c})$ indicates that commit $c$ is more likely to be relevant to issue $i$.

\subsubsection{LambdaMART}

We use LambdaMART~\cite{lambda} as the ranking model. LambdaMART is a gradient-boosted decision tree method designed for ranking tasks. Instead of treating each issue--commit pair independently, LambdaMART learns from groups of candidates associated with the same issue. This is important for our setting because the objective is not only to classify whether a pair is linked, but to order all candidate commits so that the relevant commits appear near the top of the ranked list.

LambdaMART is suitable for LinkRank for four main reasons. First, it is explicitly designed to learn rankings within query groups, which maps directly to our set-recovery formulation: each issue defines a candidate pool, and the model learns to rank commits within that issue-specific pool. Second, it optimizes ranking-oriented objectives, such as NDCG, which aligns well with the goal of placing relevant commits above irrelevant ones. Third, it can handle heterogeneous features, including cosine similarity scores, BM25 scores, temporal features, file-level features, and diff-related features, without requiring extensive feature normalization. Fourth, it is computationally efficient, as the model is trained over precomputed numerical features rather than directly over raw text or code.

\subsubsection{Hyperparameter Tuning}

We tune the LambdaMART hyperparameters using Optuna~\cite{akiba2019optuna}. For each cross-validation fold, the training portion is further split into a training set and a development set, with 20\% of the training data used for validation.

We use 50 trials with Tree-structured Parzen Estimator (TPE) sampling and tune the following hyperparameters:

\begin{itemize}
  \item number of trees: 100--1500;
  \item number of leaves per tree: 15--63;
  \item learning rate: 0.01--0.3;
  \item minimum child samples: 5--50;
  \item feature fraction: 0.5--1.0;
  \item bagging fraction: 0.5--1.0.
\end{itemize}

We implement LambdaMART using LightGBM's  \textit{lambdarank} objective~\cite{ke2017lightgbm}. LightGBM provides an efficient implementation of gradient-boosted decision trees and supports query-level grouping, which is required for learning-to-rank. In our setting, all candidate commits associated with the same issue are treated as one query group.

\subsection{Phase IV: Selection Strategies}\label{sec:selection}

After the ranking model assigns scores to all commits in an issue's candidate pool, LinkRank must construct the final predicted set of linked commits, denoted as $\widehat{\mathcal{C}}(i)$. A simple strategy would be to select the top-$K$ commits from the ranked list. However, this is not sufficient for realistic one-to-many recovery because the number of linked commits may not be known in advance. Moreover, after the highest-ranked commit is selected, the remaining candidates must be considered relative to the updated candidate pool.

To address this, LinkRank uses an iterative pick--remove--renormalize strategy. At each step, the highest-scoring candidate is selected if it satisfies the stopping criterion. The selected commit is then added to the predicted set and removed from the candidate pool. The scores of the remaining candidates are then renormalized within the reduced pool before the next selection step. This process continues until the stopping rule indicates that no further commit should be selected.

\subsubsection{Iterative Pick--Remove--Renormalize}

For each issue $i$, let $\mathcal{P}_i$ be its candidate pool and let $\mathcal{R}$ be the trained ranking model. LinkRank proceeds as follows:

\begin{enumerate}
  \item \textbf{Score:} Compute a ranking score $s(i,c) = \mathcal{R}(\mathbf{x}_{i,c})$ for each candidate commit $c \in \mathcal{P}_i$.

  \item \textbf{Normalize:} Compute min--max normalized scores within the issue-specific candidate pool:
    \[
    \tilde{s}(i,c) = \frac{s(i,c) - s_{\min}(i)}{s_{\max}(i) - s_{\min}(i)},
    \]
    where $s_{\max}(i) = \max_{c \in \mathcal{P}_i} s(i,c)$ and $s_{\min}(i) = \min_{c \in \mathcal{P}_i} s(i,c)$.

  \item \textbf{Pick:} Select the highest-scoring candidate:
    \[
    c^\star = \arg\max_{c \in \mathcal{P}_i} s(i,c).
    \]

  \item \textbf{Check stopping rule:} If $c^\star$ does not satisfy the stopping criterion, terminate the selection process for issue $i$.

  \item \textbf{Remove and update:} Add $c^\star$ to the predicted set $\widehat{\mathcal{C}}(i)$ and remove it from $\mathcal{P}_i$.

  \item \textbf{Repeat:} Recompute the normalized scores over the remaining candidates and repeat the process until the stopping rule terminates selection or the candidate pool becomes empty.
\end{enumerate}

This strategy allows LinkRank to recover multiple commits for the same issue while supporting both oracle and fully automatic settings, as summarized in Algorithm~\ref{algo_1}. In the oracle setting, the process stops after selecting the known number of linked commits. In the automatic setting, the process stops when the remaining candidates no longer satisfy a score-based threshold.

\begin{algorithm*}[htbp]
\caption{LinkRank: One-to-Many Issue--Commit Recovery}
\label{algo_1}
\begin{algorithmic}[1]
\REQUIRE Issues $\mathcal{I}$, candidate pools $\{\mathcal{P}_i\}_{i \in \mathcal{I}}$, relevance labels $\{y_{i,c}\}$, stopping rule $\in \{\text{Known-}K, \text{ABS}, \text{REL}\}$
\ENSURE Predicted link sets $\{\widehat{\mathcal{C}}(i)\}_{i \in \mathcal{I}}$

\STATE \textbf{// Phase II: Feature Extraction}
\STATE Build TF-IDF+SVD representations using four views: full, code, natural language, and title
\STATE Build BM25 scores using the same four textual views
\STATE Compute SBERT similarities using three views: full, natural language, and title
\STATE Extract metadata features for each issue--commit pair

\FORALL{$i \in \mathcal{I}$}
  \FORALL{$c \in \mathcal{P}_i$}
    \STATE $\mathbf{x}_{i,c} \gets$ [SVD features, BM25 features, SBERT features, metadata features]
  \ENDFOR
\ENDFOR

\STATE
\STATE \textbf{// Phase III: Ranking Model Training}
\STATE Tune LambdaMART hyperparameters on the development set
\STATE Train LambdaMART ranking model $\mathcal{R}$ using issue-level query groups

\STATE
\STATE \textbf{// Phase IV: Iterative Selection}
\FORALL{test issue $i \in \mathcal{I}_{\text{test}}$}
  \STATE $\widehat{\mathcal{C}}(i) \gets \emptyset$
  \STATE $\mathcal{A} \gets \mathcal{P}_i$ \hfill \textit{// active candidate pool}

  \FORALL{$c \in \mathcal{A}$}
    \STATE $s(i,c) \gets \mathcal{R}(\mathbf{x}_{i,c})$
  \ENDFOR

  \REPEAT
    \STATE $c^\star \gets \arg\max_{c \in \mathcal{A}} s(i,c)$
    \STATE Compute $\tilde{s}(i,c^\star)$ using min--max normalization over $\mathcal{A}$

    \IF{stopping rule is not satisfied for $c^\star$}
      \STATE \textbf{break}
    \ENDIF

    \STATE $\widehat{\mathcal{C}}(i) \gets \widehat{\mathcal{C}}(i) \cup \{c^\star\}$
    \STATE $\mathcal{A} \gets \mathcal{A} \setminus \{c^\star\}$
    \STATE Recompute normalized scores over the updated active pool $\mathcal{A}$
  \UNTIL{$\mathcal{A} = \emptyset$}
\ENDFOR

\STATE
\STATE \textbf{Stopping rules:}
\STATE \quad \textbf{Known-$K$:} continue while $|\widehat{\mathcal{C}}(i)| < K_i$
\STATE \quad \textbf{ABS:} continue while $\tilde{s}(i,c^\star) \geq \tau$
\STATE \quad \textbf{REL:} continue while $s(i,c^\star) \geq \gamma \cdot s_{\max}(i)$

\end{algorithmic}
\end{algorithm*}

\subsubsection{Stopping Rules}

The iterative selection process requires a stopping rule to decide when no further commits should be added to the predicted set. We evaluate three stopping rules, ranging from an oracle setting to fully automatic settings.

\paragraph{Known-$K$ (Oracle).}
In this setting, the true number of linked commits for issue $i$, denoted by $K_i = |\mathcal{C}_i|$, is assumed to be known. The selection process stops after exactly $K_i$ commits have been selected. This setting provides an upper-bound evaluation of LinkRank because it measures ranking quality without requiring the model to estimate how many commits should be selected.

\paragraph{ABS (Absolute Threshold).}
In the ABS setting, selection continues while the normalized score of the highest-ranked remaining candidate is greater than or equal to a threshold $\tau$:
\[
\tilde{s}(i,c^\star) \geq \tau.
\]
The threshold $\tau \in [0,1]$ is tuned on the development set. This rule uses a fixed score cutoff after min--max normalization and therefore tests whether the model scores can support automatic selection without knowing $K_i$ in advance.

\paragraph{REL (Relative Threshold).}
In the REL setting, selection continues while the score of the highest-ranked remaining candidate is sufficiently close to the best score in the current candidate pool:
\[
s(i,c^\star) \geq \gamma \cdot s_{\max}(i),
\]
where $\gamma \in [0,1]$ is tuned on the development set and $s_{\max}(i)$ is recomputed after each removal. This rule adapts the stopping decision to the score distribution of each issue-specific candidate pool.

\subsubsection{Threshold Tuning}

For ABS and REL, the thresholds $\tau$ and $\gamma$ are tuned independently for each cross-validation fold using the development set. We perform grid search over the interval $[0.01,0.99]$ with a step size of 0.01 and select the threshold that maximizes F1 on the development set. The selected thresholds are then applied to the corresponding test fold.

\section{Experimental Setup}

\subsection{Experimental Settings}

All experiments were conducted on a dedicated server equipped with an
\textit{NVIDIA RTX 4500 Ada Generation GPU} (24\,GB VRAM), CUDA 12.5,
driver 555.42.06, a multi-core Intel CPU, and 64\,GB RAM running
Ubuntu Linux (64-bit). Our implementation uses Python~3.10 with
\texttt{LightGBM}~\cite{ke2017lightgbm} for LambdaMART ranking,
\texttt{sentence-transformers}~\cite{reimers2019sentence} for SBERT
embeddings (\texttt{all-mpnet-base-v2}), \texttt{rank-bm25} for BM25
scoring and \texttt{Optuna}~\cite{akiba2019optuna} for hyperparameter
tuning.

To ensure reliable evaluation, we adopt 5-fold stratified
cross-validation~\cite{stone1974cross}, stratified by $K$ (the number
of linked commits per issue) so that every fold preserves the natural
$K$-distribution. Within each fold, the training set is further split
80/20 into train and development subsets; the development set is used
for Optuna hyperparameter tuning (50 trials) and threshold selection
($\tau$, $\gamma$). All reported results are mean $\pm$
standard deviation across the five test folds.

\subsection{Evaluation Metrics}

We use two groups of evaluation metrics. First, we use set-based metrics to evaluate the final predicted commit set against the ground-truth commit set for each issue. Second, we use ranking metrics to evaluate the quality of the ranked candidate list before applying the selection strategy. Together, these metrics allow us to assess both final one-to-many recovery performance and the quality of the underlying ranking.

\subsubsection{Set-Based Metrics (Precision, Recall, F1)}

In the one-to-many setting, each issue $i$ is associated with a ground-truth set of linked commits, denoted as $\mathcal{C}_i$, and the model produces a predicted set of commits, denoted as $\widehat{\mathcal{C}}(i)$. We evaluate the quality of the predicted set by computing issue-wise Precision, Recall, and F1-score. For each issue $i$, these metrics are defined as follows:

\[
\begin{aligned}
\text{Precision}^{(i)} &=
\frac{|\widehat{\mathcal{C}}(i) \cap \mathcal{C}_i|}
{|\widehat{\mathcal{C}}(i)|}, \\[0.4em]
\text{Recall}^{(i)} &=
\frac{|\widehat{\mathcal{C}}(i) \cap \mathcal{C}_i|}
{|\mathcal{C}_i|}, \\[0.4em]
\text{F1}^{(i)} &=
\frac{2 \cdot \text{Precision}^{(i)} \cdot \text{Recall}^{(i)}}
{\text{Precision}^{(i)} + \text{Recall}^{(i)}}.
\end{aligned}
\]

If $\widehat{\mathcal{C}}(i)$ is empty, we define $\text{Precision}^{(i)} = 0$. Similarly, if both precision and recall are zero, we define $\text{F1}^{(i)} = 0$. The final scores are macro-averaged across all issues so that each issue contributes equally, regardless of the number of linked commits:

\[
\text{F1} =
\frac{1}{|\mathcal{I}|}
\sum_{i \in \mathcal{I}} \text{F1}^{(i)}.
\]

These metrics are reported under three selection strategies discussed in Section~\ref{sec:selection}.

\subsubsection{Ranking Metrics}

To evaluate the quality of the ranked candidate list \emph{before}
any threshold-based selection, we report three standard information
retrieval metrics computed from the initial (non-iterative) ranking
produced by LambdaMART:

\paragraph{Mean Reciprocal Rank (MRR).}
MRR measures how early the first relevant commit appears in the ranked
list:
\[
\text{MRR} = \frac{1}{|\mathcal{I}|} \sum_{i \in \mathcal{I}}
\frac{1}{\text{rank}_i},
\]
where $\text{rank}_i$ is the position of the highest-ranked
ground-truth commit for issue $i$. MRR\,=\,1 means the top-ranked
candidate is always correct.

\paragraph{Normalized Discounted Cumulative Gain at $K$ (NDCG@$K$).}
NDCG@$K$ evaluates whether the top $K_i$ positions contain relevant
commits, with a logarithmic discount for lower positions:
\[
\text{DCG@}K_i = \sum_{j=1}^{K_i} \frac{y_{i,\pi(j)}}{\log_2(j+1)},
\quad
\text{NDCG@}K_i = \frac{\text{DCG@}K_i}{\text{IDCG@}K_i},
\]
where $\pi(j)$ is the commit at rank $j$, $y_{i,\pi(j)} \in \{0,1\}$
is its relevance label, and IDCG@$K_i$ is the ideal DCG (all relevant
commits ranked first). We set $K = K_i$ (the true number of linked
commits) for each issue, then macro-average across issues.

\paragraph{Precision at $K$ (P@$K$).}
P@$K$ measures the fraction of relevant commits among the top $K_i$
positions:
\[
\text{P@}K_i = \frac{1}{K_i} \sum_{j=1}^{K_i} y_{i,\pi(j)}.
\]

\subsubsection{Statistical Significance}

To verify that performance differences between LinkRank and each
baseline are not due to chance, we apply the one-sided \textbf{Wilcoxon
signed-rank test}~\cite{woolson2007wilcoxon} ($H_1$: LinkRank $>$
Baseline, $\alpha = 0.05$) on the per-fold Known-K F1 scores. With
$n = 5$ folds, the minimum achievable $p$-value is $p = 0.03125$
(all folds consistent). We additionally report \textbf{Cliff's
delta}~\cite{cliff1993dominance} as a non-parametric effect size:
$|\delta| < 0.147$ (negligible), $< 0.33$ (small), $< 0.474$ (medium),
$\geq 0.474$ (large).

\subsection{Baselines}\label{sec:baselines}
We compare LinkRank against four representative  issue--commit linking approaches spanning different  paradigms. EALink~\cite{ealink} is a multi-task pre-trained framework that distills CodeBERT into a compact student encoder and fine-tunes it jointly with contrastive learning (for inter-commit correlation) and an auxiliary issue-code link prediction task for issue-commit link recovery. MPLinker~\cite{mplinker} is a prompt-tuning 
approach that wraps issue--commit pairs into cloze-style 
templates and fine-tunes RoBERTa-base with PGD adversarial 
training for binary link classification. Both methods 
treat issue--commit pairs independently and do not model 
the candidate pool as a whole, which limits their ability 
to recover complete commit sets.

EasyLink~\cite{huang2025back} and 
LinkAnchor~\cite{akhavan2025linkanchor} represent more 
recent LLM-assisted approaches. EasyLink operates in two 
stages: it first encodes issues and commits using SBERT 
to retrieve an initial set of candidates via cosine 
similarity, and then reranks the retrieved candidates 
using GPT-4o to select the most relevant commits.
LinkAnchor is an LLM-agent-based approach that uses 
GPT-4o-mini with tool access to inspect repository 
context, originally designed to identify a single 
resolving commit; we adapt it to our one-to-many setting 
by prompting the model to identify all relevant commits 
from the candidate pool. Since the full candidate pool often exceeds the practical 
context window of GPT-4o-mini, reproducing LinkAnchor at 
scale is impractical due to the substantial API cost of 
reasoning over hundreds of candidate commits per issue. 
We therefore evaluate GPT-OSS-120B as a more feasible 
alternative, providing the model with the issue title, 
issue body, and a reduced candidate set capped at 80 
commits per query. While this creates a simplified setting 
relative to the full retrieval problem, it still allows 
us to assess whether open-source LLM-based reasoning can 
support complete commit-set recovery without relying on 
expensive proprietary APIs. All baselines are evaluated 
on the same $K \leq 7$ datasets with identical 5-fold 
splits, and all produce per-candidate scores or ranked 
outputs from which we derive the same evaluation metrics 
as LinkRank.

\section{Results}
\label{results}

\textit{RQ1: How does LinkRank compare against existing 
baselines for one-to-many issue--commit recovery?}
\vspace{0.5em}

Tables~\ref{tab:rq1_known} and~\ref{tab:rq1_unknown} 
present the per-project results under Known-$K$ and 
Unknown-$K$ settings, respectively. All results are 
averaged over 5-fold stratified cross-validation with 
mean $\pm$ standard deviation reported.

\begin{table*}[htbp]
\centering
\caption{Known-$K$ F1 (\%) comparison across six projects using 5-fold cross-validation. Best result per project is shown in \textbf{bold}. $\dagger$ indicates statistically significant improvement over the baseline using a one-sided Wilcoxon signed-rank test ($p < 0.05$).}
\renewcommand{\arraystretch}{1.3}
\setlength{\tabcolsep}{5pt}
\label{tab:rq1_known}
\scriptsize
\begin{tabular}{lccccccc}
\toprule
\textbf{Method} & \textbf{PyTorch} & \textbf{Dubbo} &
\textbf{Iceberg} & \textbf{Beam} & \textbf{DataFusion} &
\textbf{MXNet} & \textbf{Average} \\
\midrule
EALink
  & 13.58 {\scriptsize$\pm$4.40}
  & 13.90 {\scriptsize$\pm$3.38}
  & 28.37 {\scriptsize$\pm$3.13}
  & 20.68 {\scriptsize$\pm$4.45}
  & 22.86 {\scriptsize$\pm$3.86}
  & 11.75 {\scriptsize$\pm$3.11}
  & 18.52 \\
MPLinker
  & 16.83 {\scriptsize$\pm$5.03}
  & 23.54 {\scriptsize$\pm$12.40}
  & 34.77 {\scriptsize$\pm$3.79}
  & 25.91 {\scriptsize$\pm$6.26}
  & 21.77 {\scriptsize$\pm$10.21}
  & 20.96 {\scriptsize$\pm$5.71}
  & 23.96 \\
EasyLink
  & 26.09 {\scriptsize$\pm$3.57}
  & 32.96 {\scriptsize$\pm$1.67}
  & 53.07 {\scriptsize$\pm$5.08}
  & 34.50 {\scriptsize$\pm$3.95}
  & 38.20 {\scriptsize$\pm$1.74}
  & 30.48 {\scriptsize$\pm$4.46}
  & 35.88 \\
LinkAnchor
  & 54.30 {\scriptsize$\pm$5.37}
  & 53.79 {\scriptsize$\pm$5.21}
  & 52.67 {\scriptsize$\pm$3.36}
  & 46.98 {\scriptsize$\pm$2.18}
  & 35.86 {\scriptsize$\pm$1.70}
  & 46.73 {\scriptsize$\pm$2.71}
  & 48.39 \\
\midrule
\textbf{LinkRank (ours)}
  & \textbf{76.38}$^\dagger$ {\scriptsize$\pm$3.99}
  & \textbf{76.96}$^\dagger$ {\scriptsize$\pm$6.39}
  & \textbf{80.68}$^\dagger$ {\scriptsize$\pm$2.78}
  & \textbf{69.69}$^\dagger$ {\scriptsize$\pm$2.94}
  & \textbf{65.62}$^\dagger$ {\scriptsize$\pm$1.18}
  & \textbf{69.69}$^\dagger$ {\scriptsize$\pm$3.64}
  & \textbf{73.17} \\
\bottomrule
\end{tabular}
\end{table*}

\begin{table*}[htbp]
\centering
\caption{Unknown-$K$ F1 (\%) under ABS and REL stopping rules using 5-fold cross-validation. Best result per project is shown in \textbf{bold}.}
\renewcommand{\arraystretch}{1.3}
\setlength{\tabcolsep}{4pt}
\label{tab:rq1_unknown}
\scriptsize
\begin{tabular}{llccccccc}
\toprule
\textbf{Method} & \textbf{Rule} & \textbf{PyTorch} & \textbf{Dubbo} &
\textbf{Iceberg} & \textbf{Beam} & \textbf{DataFusion} &
\textbf{MXNet} & \textbf{Avg} \\
\midrule
\multirow{2}{*}{EALink}
  & ABS & 11.85 {\scriptsize$\pm$1.41} & 13.09 {\scriptsize$\pm$1.75}
  & 23.86 {\scriptsize$\pm$2.88} & 16.41 {\scriptsize$\pm$4.85}
  & 12.42 {\scriptsize$\pm$1.21} & 9.08 {\scriptsize$\pm$1.35} & 14.45 \\
  & REL    & 12.83 {\scriptsize$\pm$2.34} & 13.42 {\scriptsize$\pm$1.99}
  & 24.95 {\scriptsize$\pm$2.00} & 17.01 {\scriptsize$\pm$5.44}
  & 14.38 {\scriptsize$\pm$1.25} & 9.97 {\scriptsize$\pm$1.98} & 15.43 \\
\midrule
\multirow{2}{*}{MPLinker}
  & ABS & 19.81 {\scriptsize$\pm$7.16} & 23.58 {\scriptsize$\pm$8.20}
  & 41.00 {\scriptsize$\pm$4.03} & 24.67 {\scriptsize$\pm$10.63}
  & 17.45 {\scriptsize$\pm$1.07} & 24.31 {\scriptsize$\pm$12.13} & 25.13 \\
  & REL    & 18.97 {\scriptsize$\pm$7.58} & 21.81 {\scriptsize$\pm$10.77}
  & 40.85 {\scriptsize$\pm$3.99} & 24.67 {\scriptsize$\pm$10.67}
  & 17.38 {\scriptsize$\pm$1.06} & 24.05 {\scriptsize$\pm$12.05} & 24.62 \\
\midrule
\multirow{2}{*}{EasyLink}
  & ABS & 36.32 {\scriptsize$\pm$3.32} & 44.09 {\scriptsize$\pm$2.07}
  & 60.43 {\scriptsize$\pm$3.68} & 42.95 {\scriptsize$\pm$3.27}
  & 45.98 {\scriptsize$\pm$2.17} & 38.49 {\scriptsize$\pm$4.56} & 44.71 \\
  & REL    & 36.60 {\scriptsize$\pm$3.29} & 44.43 {\scriptsize$\pm$2.08}
  & 61.21 {\scriptsize$\pm$3.59} & 43.72 {\scriptsize$\pm$2.96}
  & 46.72 {\scriptsize$\pm$2.19} & 39.41 {\scriptsize$\pm$4.60} & 45.35 \\
\midrule
\multirow{2}{*}{LinkAnchor}
  & ABS & 62.82 {\scriptsize$\pm$6.68} & 59.81 {\scriptsize$\pm$5.88}
  & 65.34 {\scriptsize$\pm$4.04} & 59.48 {\scriptsize$\pm$1.95}
  & 55.06 {\scriptsize$\pm$1.20} & 52.19 {\scriptsize$\pm$4.24} & 59.12 \\
  & REL    & 48.34 {\scriptsize$\pm$5.16} & 39.16 {\scriptsize$\pm$7.43}
  & 44.68 {\scriptsize$\pm$4.49} & 41.64 {\scriptsize$\pm$5.25}
  & 38.92 {\scriptsize$\pm$1.14} & 31.58 {\scriptsize$\pm$1.52} & 40.72 \\
\midrule
\multirow{2}{*}{\textbf{LinkRank (ours)}}
  & ABS & \textbf{70.88} {\scriptsize$\pm$3.68}
  & \textbf{68.53} {\scriptsize$\pm$3.73}
  & \textbf{71.14} {\scriptsize$\pm$2.80}
  & \textbf{62.50} {\scriptsize$\pm$4.00}
  & \textbf{59.21} {\scriptsize$\pm$1.28}
  & \textbf{63.37} {\scriptsize$\pm$3.23}
  & \textbf{65.94} \\
  & REL    & \textbf{70.25} {\scriptsize$\pm$1.49}
  & \textbf{67.36} {\scriptsize$\pm$3.85}
  & \textbf{66.78} {\scriptsize$\pm$2.33}
  & \textbf{60.57} {\scriptsize$\pm$4.26}
  & \textbf{46.90} {\scriptsize$\pm$6.37}
  & \textbf{61.65} {\scriptsize$\pm$2.19}
  & \textbf{62.19} \\
\bottomrule
\end{tabular}
\end{table*}

\paragraph{Analysis.}
The results show that LinkRank consistently outperforms 
all baselines across the six evaluated projects. In the 
Known-$K$ setting, where the true number of linked commits 
is assumed to be known, LinkRank achieves an average F1 
score of 73.17\%. The strongest baseline is LinkAnchor, 
which achieves 48.39\% F1. This shows that LinkRank 
remains substantially more effective at selecting the 
complete set of relevant commits when multiple commits 
may be associated with the same issue. The same trend is 
observed in the Unknown-$K$ setting, where the number of 
linked commits must be inferred automatically. LinkRank 
achieves 65.94\% F1 with ABS and 62.19\% F1 with REL; 
under ABS, the strongest baseline is LinkAnchor at 
59.12\% F1, while under REL, the strongest baseline is 
EasyLink at 45.35\% F1. The gap between Known-$K$ and 
Unknown-$K$ performance remains moderate for LinkRank, 
suggesting that the proposed stopping rules recover much 
of the benefit of the oracle setting while operating 
without prior knowledge of the number of linked commits. 
The ABS setting performs slightly better than REL, 
indicating that the min--max normalised threshold provides 
a more effective stopping criterion in this evaluation.

\paragraph{Baseline comparison.}
The weaker performance of the baselines can be explained 
by their original task formulations. EALink is designed 
primarily as a pairwise link prediction model and is not 
optimised for selecting commits from a large candidate 
pool under severe class imbalance. MPLinker also treats 
issue--commit pairs independently, which limits its 
ability to compare candidates within the same 
issue-specific pool. EasyLink performs better than EALink 
and MPLinker because semantic similarity is useful for 
retrieving related commits, but it relies mainly on 
embedding-based similarity and does not combine 
complementary signals such as temporal proximity, 
file-level overlap, and diff characteristics. LinkAnchor 
is more competitive than the other baselines, particularly 
under the Known-$K$ and ABS settings, because it uses an 
LLM-based workflow with tool access to reason over 
repository context. However, even LinkAnchor remains 
substantially below LinkRank, suggesting that the 
performance gap is not explained by model capacity alone 
but by problem formulation: methods designed for pairwise 
decisions or single-commit retrieval cannot naturally 
compare candidates within the same issue-specific pool 
or decide when the recovered commit set is complete.

\paragraph{Statistical significance.}
To further validate the observed improvements, we compare 
LinkRank against each baseline using a one-sided Wilcoxon 
signed-rank test on the per-fold Known-$K$ F1 scores. 
LinkRank's improvements are statistically significant 
across all baseline comparisons, with $p = 0.031$ in each 
case, the minimum achievable $p$-value with five folds, indicating that LinkRank outperforms each baseline 
consistently across all folds. Cliff's $\delta = 1.00$ 
further confirms a large effect size, showing that no 
baseline fold score exceeds the corresponding LinkRank 
fold score, and that the performance gains reflect a 
consistent and practically meaningful improvement over 
existing methods.

\textit{RQ2: How well does LinkRank rank relevant commits 
before the final selection step?}
\vspace{0.5em}

While RQ1 evaluates the final recovered set of commits, 
we also examine the quality of the ranked candidate list 
before applying the iterative selection strategy. This is 
important because a one-to-many recovery method should 
not only select a final set of commits, but also place 
relevant commits near the top of the candidate list. 
Table~\ref{tab:rq2_ranking} reports three ranking metrics: 
MRR, NDCG@$K$, and P@$K$.

\paragraph{Analysis.}
The results show that LinkRank achieves the best ranking 
performance across all six projects. LinkRank obtains an 
average MRR of 87.19\%, compared with 57.49\% for the 
strongest baseline, EasyLink, indicating that LinkRank 
consistently places at least one relevant commit much 
closer to the top of the ranked list. The NDCG@$K$ and 
P@$K$ results further confirm that LinkRank is effective 
at ranking multiple relevant commits near the top 
positions, achieving an average NDCG@$K$ of 74.87\% and 
P@$K$ of 72.74\%, compared with 48.54\% and 48.39\% for 
LinkAnchor. These results are particularly important in 
the one-to-many setting, where a useful model must rank 
several relevant commits highly rather than retrieving 
only one, and they suggest that the learning-to-rank 
formulation provides a strong and well-ordered basis for 
the subsequent iterative selection step.

\begin{table*}[htbp]
\centering
\caption{Ranking quality metrics (\%) computed from the initial ranked list across six projects using 5-fold cross-validation. Best result per project is shown in \textbf{bold}.}
\renewcommand{\arraystretch}{1.3}
\setlength{\tabcolsep}{5pt}
\label{tab:rq2_ranking}
\scriptsize
\begin{tabular}{llccccccc}
\toprule
\textbf{Method} & \textbf{Metric} & \textbf{PyTorch} & \textbf{Dubbo} &
\textbf{Iceberg} & \textbf{Beam} & \textbf{DataFusion} &
\textbf{MXNet} & \textbf{Avg} \\
\midrule
\multirow{3}{*}{EALink}
  & MRR    & 22.74 & 23.80 & 44.60 & 33.28 & 39.79 & 21.28 & 30.92 \\
  & NDCG@K & 13.97 & 14.14 & 29.65 & 21.32 & 24.48 & 12.18 & 19.29 \\
  & P@K    & 13.58 & 13.90 & 28.37 & 20.68 & 22.86 & 11.75 & 18.52 \\
\midrule
\multirow{3}{*}{MPLinker}
  & MRR    & 28.21 & 35.96 & 58.07 & 42.20 & 40.23 & 37.35 & 40.34 \\
  & NDCG@K & 17.50 & 24.40 & 36.97 & 27.43 & 23.55 & 22.26 & 25.35 \\
  & P@K    & 16.83 & 23.54 & 34.77 & 25.91 & 21.77 & 20.96 & 23.96 \\
\midrule
\multirow{3}{*}{EasyLink}
  & MRR    & 43.22 & 49.18 & 78.59 & 55.99 & 70.26 & 47.72 & 57.49 \\
  & NDCG@K & 27.93 & 34.40 & 57.09 & 37.37 & 43.46 & 32.17 & 38.74 \\
  & P@K    & 26.09 & 32.96 & 53.07 & 34.50 & 38.20 & 30.48 & 35.88 \\
\midrule
\multirow{3}{*}{LinkAnchor}
  & MRR    & 56.40 & 55.41 & 53.84 & 50.91 & 37.85 & 50.06 & 50.75 \\
  & NDCG@K & 54.28 & 53.84 & 52.57 & 47.45 & 36.07 & 47.02 & 48.54 \\
  & P@K    & 54.30 & 53.79 & 52.67 & 46.98 & 35.86 & 46.73 & 48.39 \\
\midrule
\multirow{3}{*}{\textbf{LinkRank (ours)}}
  & MRR    & \textbf{85.54} & \textbf{88.01} & \textbf{92.08} & \textbf{84.70} & \textbf{89.53} & \textbf{83.30} & \textbf{87.19} \\
  & NDCG@K & \textbf{76.60} & \textbf{77.70} & \textbf{81.59} & \textbf{71.55} & \textbf{71.31} & \textbf{70.48} & \textbf{74.87} \\
  & P@K    & \textbf{75.56} & \textbf{76.29} & \textbf{79.63} & \textbf{69.19} & \textbf{67.10} & \textbf{68.67} & \textbf{72.74} \\
\bottomrule
\end{tabular}
\end{table*}

\textit{RQ3: What is the contribution of each feature 
group to LinkRank's performance?}
\vspace{0.5em}

To understand the contribution of different feature 
groups, we conduct an ablation study by training 
LinkRank with different combinations of lexical, 
retrieval-based, semantic, and metadata features. 
Table~\ref{tab:ablation} reports the Known-$K$ F1 
scores on the PyTorch dataset.

\begin{table}[htbp]
\centering
\caption{Feature ablation study on the PyTorch dataset using Known-$K$ F1 (\%). Each row uses only the listed feature groups.}
\renewcommand{\arraystretch}{1.2}
\label{tab:ablation}
\scriptsize
\begin{tabular}{clcc}
\toprule
\textbf{\#} & \textbf{Configuration} & \textbf{Features} & \textbf{Known-$K$ F1} \\
\midrule
\multicolumn{4}{l}{\textit{Single feature groups}} \\
1  & SBERT only          & 3  & 44.69 \\
2  & SVD only            & 4  & 25.99 \\
3  & BM25 only           & 4  & 37.85 \\
4  & Metadata only       & 6  & 38.73 \\
\midrule
\multicolumn{4}{l}{\textit{Pairwise combinations with SBERT}} \\
5  & SBERT + Metadata    & 9  & 64.84 \\
6  & SBERT + SVD         & 7  & 44.93 \\
7  & SBERT + BM25        & 7  & 49.59 \\
\midrule
\multicolumn{4}{l}{\textit{Three feature groups}} \\
8  & SBERT + SVD + Metadata       & 13 & 65.90 \\
9  & SBERT + BM25 + Metadata      & 13 & 71.22 \\
10 & SVD + BM25 + Metadata        & 14 & 62.74 \\
11 & SBERT + SVD + BM25           & 11 & 49.79 \\
\midrule
\multicolumn{4}{l}{\textit{All feature groups}} \\
12 & SBERT + SVD + BM25 + Metadata & 17 & \textbf{76.38} \\
\bottomrule
\end{tabular}
\end{table}

\paragraph{Analysis.}
The ablation results reveal three consistent trends 
that together explain why LinkRank's multi-view 
feature design is important for set-recovery. First, 
no single feature group is sufficient by itself. 
Individual feature groups achieve between 25.99\% 
and 44.69\% F1. SBERT is the strongest single group, 
suggesting that semantic similarity is useful for 
issue--commit recovery. However, its performance 
remains substantially below the best multi-feature 
configurations, showing that semantic similarity 
alone cannot fully capture one-to-many traceability 
links.

Second, metadata features provide a strong 
complementary signal. When metadata is removed from 
the combination of SBERT, SVD, and BM25, the F1 
score is 49.79\%. Adding metadata increases the 
score to 76.38\%, an improvement of 26.59 points. 
This indicates that temporal proximity, file-level 
overlap, and change-size information capture 
important context unavailable from textual similarity 
alone. For example, two commits may use similar 
terms, but the commit that occurs closer to the 
issue resolution period or modifies files mentioned 
in the issue may be more likely to be relevant.

Third, the best performance is achieved when all 
four feature groups are combined. The full 
configuration achieves the highest Known-$K$ F1 
of 76.38\%, suggesting that the feature groups 
capture orthogonal aspects of relevance: SBERT 
captures semantic relatedness, BM25 captures 
direct retrieval-based matching, SVD captures 
latent lexical structure, and metadata captures 
temporal and structural repository context. Although 
the SBERT + BM25 + Metadata configuration already 
performs strongly at 71.22\% F1, adding SVD further 
improves performance, showing that lexical 
latent-space features still contribute useful 
additional information. We observed a similar trend 
across the other datasets, where combining multiple 
feature groups consistently outperformed any single 
group alone. Notably, the largest single gain comes 
from adding metadata rather than from switching 
between textual representations, suggesting that 
temporal and structural context, information 
unavailable to purely text-based classifiers, is 
particularly valuable for one-to-many issue--commit 
recovery.

\textit{RQ4: What are the training and inference costs of 
LinkRank compared to baselines?}
\vspace{0.5em}

\begin{figure}
  \centering
  \includegraphics[width=0.99\linewidth]{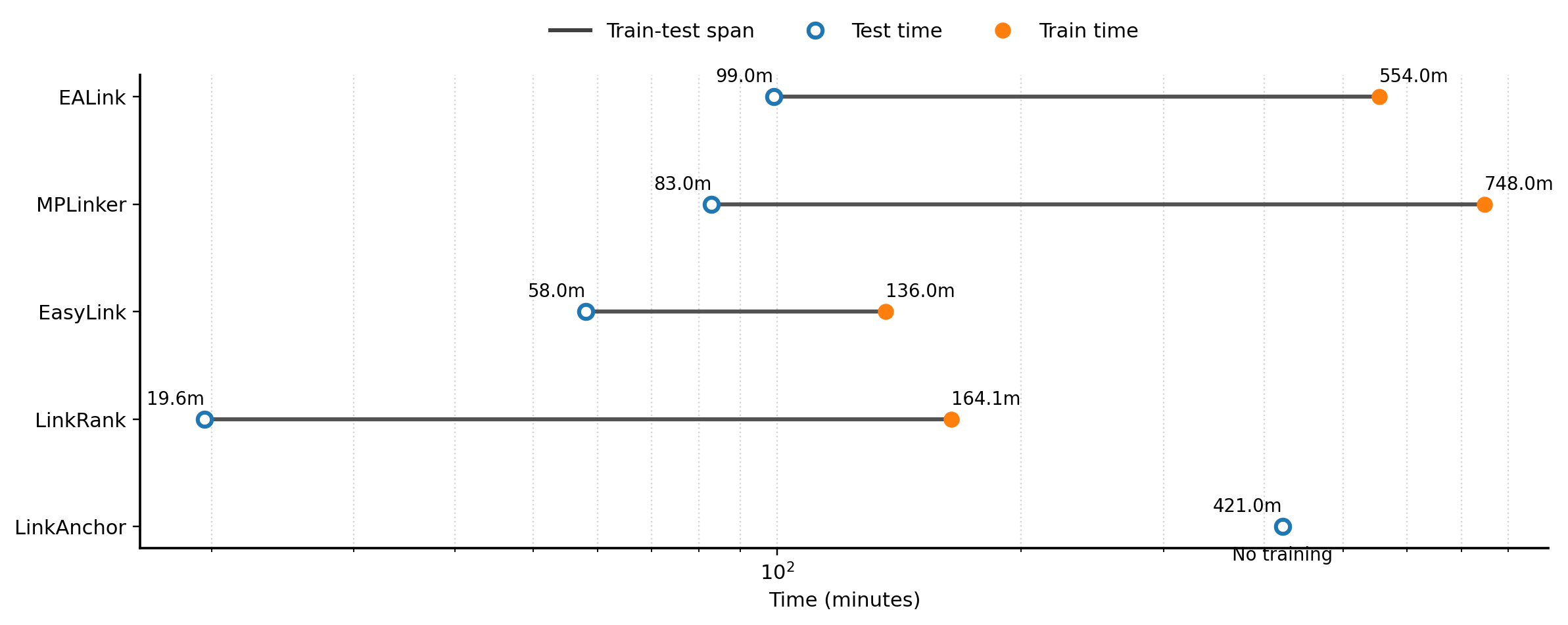}
  \caption{Training and test-time comparison of LinkRank and baseline methods. The blue hollow marker shows test/inference time, the orange marker shows training time, and the connecting line indicates the span between the two.}

  \label{fig:time_efficiency}
\end{figure}

Figure~\ref{fig:time_efficiency} compares the training 
and test-time costs of LinkRank with the baseline methods. 
The reported times are measured in minutes over the full 
evaluation pipeline. For methods that require a training 
phase, the left marker denotes test/inference time and 
the right marker denotes training time. LinkAnchor does 
not require task-specific training, so only its 
test/inference time is shown. The connecting line 
indicates the span between test and training costs for 
methods with both stages.

\paragraph{Analysis.}
The results show that LinkRank provides a favorable balance between training and inference cost. LinkRank 
has the lowest test/inference time among all compared 
methods at 19.6 minutes, compared with 58.0 minutes for 
EasyLink, 83.0 minutes for MPLinker, 99.0 minutes for 
EALink, and 421.0 minutes for LinkAnchor. Although 
LinkAnchor does not require task-specific training, its 
inference cost is substantially higher because it relies 
on repeated LLM-based reasoning over candidate commits, demonstrating that avoiding training does not 
necessarily lead to lower overall runtime when inference 
itself is expensive. In terms of training time, LinkRank 
requires 164.1 minutes, slightly higher than EasyLink 
at 136.0 minutes but substantially lower than EALink 
and MPLinker at 554.0 and 748.0 minutes respectively. 
This relatively low training cost stems from the fact 
that LambdaMART is trained over precomputed numerical 
features rather than directly over raw text or code; 
feature extraction, including SBERT encoding, BM25 
scoring, and TF-IDF+SVD representation, accounts for 
most of the computational overhead, while the ranking 
model itself remains lightweight. Compared with neural 
baselines, LinkRank avoids expensive graph neural network 
training or transformer fine-tuning, and compared with 
LinkAnchor, it avoids repeated LLM-based reasoning at 
inference time, making it well suited for large-scale 
repository-level traceability recovery where many 
candidate commits must be scored per issue and repeated 
LLM-based reasoning is impractical.

\section{Discussion}\label{sec:discussion}

In this section, we first discuss the stability of LinkRank's automatic stopping rules through threshold sensitivity analysis. We then summarize the main limitations and threats to validity of our study.

\subsection{Threshold Sensitivity}
\label{sec:sensitivity_discussion}

\begin{figure}
  \centering
  \includegraphics[width=0.99\linewidth]{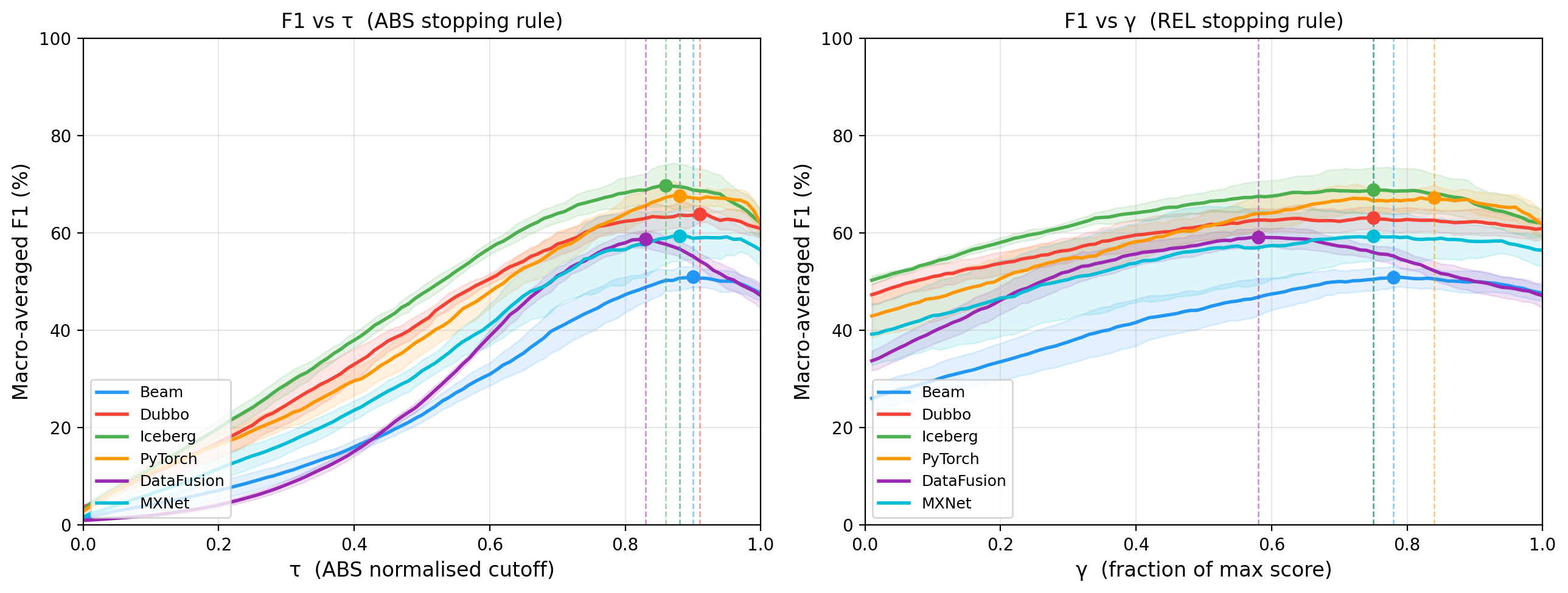}
  \caption{Sensitivity of LinkRank to stopping thresholds $\tau$ (ABS) and $\gamma$ (REL). Lines show mean F1 over 5-fold cross-validation, bands show $\pm$1 standard deviation, and dots indicate development-tuned thresholds. The broad plateau shows that LinkRank is robust to threshold choice.}
  \label{sensitivity}
\end{figure}

While the Known-$K$ setting evaluates the ranking model when the number of linked commits is available, the Unknown-$K$ setting is more realistic because the model must decide when to stop selecting commits. Therefore, the practical usefulness of LinkRank depends not only on its ranking quality, but also on the stability of its automatic stopping rules. To examine this aspect, we analyze the sensitivity of the ABS and REL stopping rules to their threshold values $\tau$ and $\gamma$, respectively. These thresholds directly affect the number of commits returned for each issue, so an important question is whether LinkRank depends on carefully tuned threshold values. To study this, we sweep $\tau \in [0,1]$ for ABS and $\gamma \in [0.01,1]$ for REL using 100 equally spaced values for each of the six datasets. For every threshold value, we apply the corresponding stopping rule to the ranked candidate lists produced by LinkRank and compute the macro-averaged F1 score across test issues under 5-fold cross-validation. As shown in Figure~\ref{sensitivity}, LinkRank is not dependent on a single narrow threshold value. For ABS, performance increases steadily as $\tau$ increases from low values and then remains stable across a broad high-threshold region, with the best-performing thresholds concentrated around $0.8$--$0.9$. REL shows a similar pattern, although its optimal $\gamma$ values vary more across datasets. In both cases, the F1 curves remain close to their peak over a practical range of threshold values, and the shaded regions indicate moderate fold-level variance around the peak regions. This suggests that the Unknown-$K$ results are not driven by a carefully chosen threshold, but reflect stable behavior across cross-validation splits. Overall, the sensitivity analysis shows that LinkRank can be applied with fixed or lightly tuned stopping thresholds in practical settings, with ABS appearing slightly more stable than REL because its optimal thresholds are more tightly concentrated across datasets.

\subsection{Limitations, Threats to Validity and Future Work}
\label{sec:threats}

We acknowledge that our study is subject to several threats to validity and these threats are discussed below, along with the steps taken to mitigate them.

\textbf{Candidate pool construction.}
We construct candidate pools using a fixed temporal window around each issue, which keeps the evaluation realistic while avoiding artificially balanced positive-negative ratios. However, the window size may affect the results: a narrow window may miss valid commits, while a wide window may introduce more irrelevant candidates. We mitigate this by choosing the window based on the observed issue--commit time-gap distribution across all six projects and verifying that more than 98\% of ground-truth commits fall within it.

\textbf{Project selection.}
We evaluate on six open-source GitHub projects spanning three programming
languages (Python: PyTorch, Apache Beam, Apache DataFusion; Java: Apache
Dubbo, Apache Iceberg; C++: MXNet) and multiple domains (deep learning
frameworks, data processing, RPC middleware).
Despite this diversity, all projects are large, actively maintained, and
hosted on GitHub.
The results may not generalize to smaller projects, closed-source codebases,
projects with sparse commit messages, or platforms with different issue-tracking
conventions (e.g., Jira, Bugzilla).

\textbf{K-distribution generalizability.}
We restrict our study to issues with $K \leq 7$ linked commits, which covers
the vast majority of real-world issues (the long tail of $K > 7$ accounts for
$<$2\% of issues in our datasets).
Issues requiring more than seven commits may exhibit different patterns that
our current feature set does not capture well; however, they are sufficiently
rare that excluding them has negligible impact on practical deployment.

\paragraph{Future work.}
We plan to investigate how large language models can be used more effectively for one-to-many issue--commit recovery. Although LinkRank outperforms existing LLM-based baselines in our evaluation, these baselines are still largely designed for one-to-one issue--commit recovery, and the results show that applying LLMs directly to complete one-to-many commit-set recovery remains challenging. A single issue may have to be matched against hundreds or even more than a thousand candidate commits, which creates practical limitations due to context-window size and the cost of repeated API calls. Therefore, an important direction is to develop better candidate filtering mechanisms that can reduce the search space before applying an LLM. After narrowing the candidate pool, locally deployable open-source LLMs could be used to reason over the remaining commits without relying on expensive proprietary APIs. Another promising direction is to use LLMs not only for prediction, but also for explanation. For each predicted link, an LLM-based component could generate a rationale explaining why a commit is relevant or irrelevant to an issue. This would make one-to-many issue--commit recovery more explainable and interpretable, and could help developers better understand the recovered traceability links.

\section{Conclusion}\label{sec:conclusion}
This paper addressed a fundamental limitation of existing 
issue--commit traceability methods: their assumption that 
each issue is resolved by a single commit. We argued that 
one-to-many traceability is better understood as an 
issue-level set-recovery problem rather than independent 
pairwise classification, and operationalised this 
reframing through \textit{LinkRank}, a learning-to-rank 
framework that models each issue as a query over a pool 
of candidate commits and recovers the complete set of 
linked commits through iterative selection. To support 
realistic evaluation, we constructed a dataset from six 
open-source GitHub repositories preserving issue-level 
commit multiplicity and reflecting the severe class 
imbalance found in real repositories. The empirical 
results consistently support the central claim: LinkRank 
outperforms all baselines under both Known-$K$ and 
Unknown-$K$ settings by a substantial margin, including 
recent LLM-based approaches with far greater model 
capacity, with ablation results further confirming that 
combining textual, semantic, retrieval-based, temporal, 
and file-level features is essential for accurate 
recovery. The central finding is that correctly framing the task as set recovery
drives most of the gain: once candidates are ranked jointly within an
issue-specific pool and selected iteratively, a lightweight ranker
consistently outperforms heavier neural and LLM-based methods designed
for one-to-one link recovery.

\section*{ CRediT authorship contribution statement}
\textbf{Abhishek Kumar:} Writing – review \& editing, Writing – original draft, Validation, Methodology, Data curation, Conceptualization. \textbf{Tuhin Mondal}: Writing – review \& editing, Methodology, Data curation. \textbf{Partha Pratim Das}: Validation, Supervision, Methodology. \textbf{Partha Pratim Chakrabarti}: Writing – review \& editing, Validation, Supervision.

\section*{ Declaration of generative AI and AI-assisted technologies in the manuscript preparation process}
The author(s) used Chatgpt to improve readability and grammar. The author(s) are fully responsible for the publication's content after using this tool/service, including reviewing and editing it as required.

\section*{Declaration of competing interest}
The authors declare that they have no known competing financial interests or personal relationships that could have appeared to influence the work reported in this paper.

\section*{Data availability}

The data is provided in our Github repository\cite{abhi_repo}.




\bibliographystyle{cas-model2-names}

\bibliography{cas-refs}

@article{r1,
  title={Bugzilla, ITracker, and other bug trackers},
  author={Serrano, Nicolas and Ciordia, Ismael},
  journal={IEEE software},
  volume={22},
  number={2},
  pages={11--13},
  year={2005},
  publisher={IEEE}
}

@inproceedings{r2,
  title={The product backlog},
  author={Sedano, Todd and Ralph, Paul and P{\'e}raire, C{\'e}cile},
  booktitle={2019 IEEE/ACM 41st International Conference on Software Engineering (ICSE)},
  pages={200--211},
  year={2019},
  organization={IEEE}
}

@book{r3,
  title={Version Control with Git: Powerful tools and techniques for collaborative software development},
  author={Loeliger, Jon and McCullough, Matthew},
  year={2012},
  publisher={" O'Reilly Media, Inc."}
}

@article{r4,
  title={Version control system: A review},
  author={Zolkifli, Nazatul Nurlisa and Ngah, Amir and Deraman, Aziz},
  journal={Procedia Computer Science},
  volume={135},
  pages={408--415},
  year={2018},
  publisher={Elsevier}
}

@inproceedings{r5,
  title={Software process data quality and characteristics: a historical view on open and closed source projects},
  author={Bachmann, Adrian and Bernstein, Abraham},
  booktitle={Proceedings of the joint international and annual ERCIM workshops on Principles of software evolution (IWPSE) and software evolution (Evol) workshops},
  pages={119--128},
  year={2009}
}

@article{r56,
  title={Frlink: Improving the recovery of missing issue-commit links by revisiting file relevance},
  author={Sun, Yan and Wang, Qing and Yang, Ye},
  journal={Information and Software Technology},
  volume={84},
  pages={33--47},
  year={2017},
  publisher={Elsevier}
}

@inproceedings{r23,
  title={Linkster: enabling efficient manual inspection and annotation of mined data},
  author={Bird, Christian and Bachmann, Adrian and Rahman, Foyzur and Bernstein, Abraham},
  booktitle={Proceedings of the eighteenth ACM SIGSOFT international symposium on Foundations of software engineering},
  pages={369--370},
  year={2010}
}

@inproceedings{r6,
  title={Relink: recovering links between bugs and changes},
  author={Wu, Rongxin and Zhang, Hongyu and Kim, Sunghun and Cheung, Shing-Chi},
  booktitle={Proceedings of the 19th ACM SIGSOFT symposium and the 13th European conference on Foundations of software engineering},
  pages={15--25},
  year={2011}
}

@inproceedings{r48,
  title={Multi-layered approach for recovering links between bug reports and fixes},
  author={Nguyen, Anh Tuan and Nguyen, Tung Thanh and Nguyen, Hoan Anh and Nguyen, Tien N},
  booktitle={Proceedings of the ACM SIGSOFT 20th International Symposium on the Foundations of Software Engineering},
  pages={1--11},
  year={2012}
}

@inproceedings{r27,
  title={On automatically generating commit messages via summarization of source code changes},
  author={Cort{\'e}s-Coy, Luis Fernando and Linares-V{\'a}squez, Mario and Aponte, Jairo and Poshyvanyk, Denys},
  booktitle={2014 IEEE 14th International Working Conference on Source Code Analysis and Manipulation},
  pages={275--284},
  year={2014},
  organization={IEEE}
}

@inproceedings{r28,
  title={Changescribe: A tool for automatically generating commit messages},
  author={Linares-V{\'a}squez, Mario and Cort{\'e}s-Coy, Luis Fernando and Aponte, Jairo and Poshyvanyk, Denys},
  booktitle={2015 IEEE/ACM 37th IEEE International Conference on Software Engineering},
  volume={2},
  pages={709--712},
  year={2015},
  organization={IEEE}
}

@article{a,
  title={Stochastic gradient boosting},
  author={Friedman, Jerome H},
  journal={Computational statistics \& data analysis},
  volume={38},
  number={4},
  pages={367--378},
  year={2002},
  publisher={Elsevier}
}

@article{c,
  title={Random forests},
  author={Breiman, Leo},
  journal={Machine learning},
  volume={45},
  number={1},
  pages={5--32},
  year={2001},
  publisher={Springer}
}

@inproceedings{d,
  title={Xgboost: A scalable tree boosting system},
  author={Chen, Tianqi and Guestrin, Carlos},
  booktitle={Proceedings of the 22nd acm sigkdd international conference on knowledge discovery and data mining},
  pages={785--794},
  year={2016}
}

@inproceedings{q1,
  title={Deeplink: A code knowledge graph based deep learning approach for issue-commit link recovery},
  author={Xie, Rui and Chen, Long and Ye, Wei and Li, Zhiyu and Hu, Tianxiang and Du, Dongdong and Zhang, Shikun},
  booktitle={2019 IEEE 26th International Conference on Software Analysis, Evolution and Reengineering (SANER)},
  pages={434--444},
  year={2019},
  organization={IEEE}
}

@inproceedings{q2,
  title={Automated Recovery of Issue-Commit Links Leveraging Both Textual and Non-textual Data},
  author={Mazrae, Pooya Rostami and Izadi, Maliheh and Heydarnoori, Abbas},
  booktitle={2021 IEEE International Conference on Software Maintenance and Evolution (ICSME)},
  pages={263--273},
  year={2021},
  organization={IEEE}
}

@inproceedings{q3,
  title={Rclinker: Automated linking of issue reports and commits leveraging rich contextual information},
  author={Le, Tien-Duy B and Linares-V{\'a}squez, Mario and Lo, David and Poshyvanyk, Denys},
  booktitle={2015 IEEE 23rd International Conference on Program Comprehension},
  pages={36--47},
  year={2015},
  organization={IEEE}
}

@inproceedings{q4,
  title={Improving missing issue-commit link recovery using positive and unlabeled data},
  author={Sun, Yan and Chen, Celia and Wang, Qing and Boehm, Barry},
  booktitle={2017 32nd IEEE/ACM International Conference on Automated Software Engineering (ASE)},
  pages={147--152},
  year={2017},
  organization={IEEE}
}

@inproceedings{traceability,
  title={Traceability transformed: Generating more accurate links with pre-trained bert models},
  author={Lin, Jinfeng and Liu, Yalin and Zeng, Qingkai and Jiang, Meng and Cleland-Huang, Jane},
  booktitle={2021 IEEE/ACM 43rd International Conference on Software Engineering (ICSE)},
  pages={324--335},
  year={2021},
  organization={IEEE}
}

@article{btlink,
  title={BTLink: automatic link recovery between issues and commits based on pre-trained BERT model},
  author={Lan, Jinpeng and Gong, Lina and Zhang, Jingxuan and Zhang, Haoxiang},
  journal={Empirical Software Engineering},
  volume={28},
  number={4},
  pages={1--55},
  year={2023},
  publisher={Springer}
}

@article{re1,
  title={LinkFormer: Automatic Contextualised Link Recovery of Software Artifacts in both Project-based and Transfer Learning Settings},
  author={Izadi, Maliheh and Mazrae, Pooya Rostami and Mens, Tom and van Deursen, Arie},
  journal={arXiv preprint arXiv:2211.00381},
  year={2022}
}

@inproceedings{rene1,
  title={Discovering loners and phantoms in commit and issue data},
  author={Schermann, Gerald and Brandtner, Martin and Panichella, Sebastiano and Leitner, Philipp and Gall, Harald},
  booktitle={2015 IEEE 23rd International Conference on Program Comprehension},
  pages={4--14},
  year={2015},
  organization={IEEE}
}

@inproceedings{rene2,
  title={Traceability in the wild: automatically augmenting incomplete trace links},
  author={Rath, Michael and Rendall, Jacob and Guo, Jin LC and Cleland-Huang, Jane and M{\"a}der, Patrick},
  booktitle={Proceedings of the 40th International Conference on Software Engineering},
  pages={834--845},
  year={2018}
}

@article{rene3,
  title={DeepLink: Recovering issue-commit links based on deep learning},
  author={Ruan, Hang and Chen, Bihuan and Peng, Xin and Zhao, Wenyun},
  journal={Journal of Systems and Software},
  volume={158},
  pages={110406},
  year={2019},
  publisher={Elsevier}
}

@inproceedings{ealink,
  title={EALink: An efficient and accurate pre-trained framework for issue-commit link recovery},
  author={Zhang, Chenyuan and Wang, Yanlin and Wei, Zhao and Xu, Yong and Wang, Juhong and Li, Hui and Ji, Rongrong},
  booktitle={2023 38th IEEE/ACM International Conference on Automated Software Engineering (ASE)},
  pages={217--229},
  year={2023},
  organization={IEEE}
}

@inproceedings{hermes,
  title={Hermes: Using commit-issue linking to detect vulnerability-fixing commits},
  author={Nguyen-Truong, Giang and Kang, Hong Jin and Lo, David and Sharma, Abhishek and Santosa, Andrew E and Sharma, Asankhaya and Ang, Ming Yi},
  booktitle={2022 IEEE International Conference on Software Analysis, Evolution and Reengineering (SANER)},
  pages={51--62},
  year={2022},
  organization={IEEE}
}

@inproceedings{impact_analysis,
  title={A literature review of automatic traceability links recovery for software change impact analysis},
  author={Aung, Thazin Win Win and Huo, Huan and Sui, Yulei},
  booktitle={Proceedings of the 28th International Conference on Program Comprehension},
  pages={14--24},
  year={2020}
}

@inproceedings{project,
  title={Successful deployment of requirements traceability in a commercial engineering organization... really},
  author={Panis, Michael C},
  booktitle={2010 18th IEEE International Requirements Engineering Conference},
  pages={303--307},
  year={2010},
  organization={IEEE}
}

@inproceedings{intro_testing,
  title={Using traceability to support model-based regression testing},
  author={Naslavsky, Leila and Richardson, Debra J},
  booktitle={Proceedings of the 22nd IEEE/ACM International Conference on Automated Software Engineering},
  pages={567--570},
  year={2007}
}

@inproceedings{cost1,
  title={The missing links: bugs and bug-fix commits},
  author={Bachmann, Adrian and Bird, Christian and Rahman, Foyzur and Devanbu, Premkumar and Bernstein, Abraham},
  booktitle={Proceedings of the eighteenth ACM SIGSOFT international symposium on Foundations of software engineering},
  pages={97--106},
  year={2010}
}

@inproceedings{cost3,
  title={A semantic relatedness approach for traceability link recovery},
  author={Mahmoud, Anas and Niu, Nan and Xu, Songhua},
  booktitle={2012 20th IEEE international conference on program comprehension (ICPC)},
  pages={183--192},
  year={2012},
  organization={IEEE}
}

@misc{GitHubAPI,
  author    = {GitHub},
  title     = {Rate Limits for the REST API},
  year      = {2022},
  url       = {https://docs.github.com/en/rest/using-the-rest-api/rate-limits-for-the-rest-api?apiVersion=2022-11-28},
  note      = {Accessed: February 7, 2025}
}

@article{pi_issue_pr,
  title={Pi-link: A ground-truth dataset of links between pull-requests and issues in github},
  author={Alshara, Zakarea and Shatnawi, Anas and Eyal-Salman, Hamzeh and Seriai, Abdelhak-Djamel and Shatnawi, Maad},
  journal={IEEE Access},
  volume={11},
  pages={697--710},
  year={2022},
  publisher={IEEE}
}

@book{safety,
  author    = {L. Rierson},
  title     = {Developing Safety-Critical Software: A Practical Guide for Aviation Software and DO-178C Compliance},
  publisher = {CRC Press},
  year      = {2013}
}

@article{mplinker,
  title={MPLinker: Multi-template Prompt-tuning with adversarial training for issue-commit Link recovery},
  author={Wang, Bangchao and Deng, Yang and Luo, Ruiqi and Liang, Peng and Bi, Tingting},
  journal={Journal of Systems and Software},
  pages={112351},
  year={2025},
  publisher={Elsevier}
}

@article{bm25,
  title={The probabilistic relevance framework: BM25 and beyond},
  author={Robertson, Stephen and Zaragoza, Hugo and others},
  journal={Foundations and Trends{\textregistered} in Information Retrieval},
  volume={3},
  number={4},
  pages={333--389},
  year={2009},
  publisher={Now Publishers, Inc.}
}

@article{lambda,
  title={From RankNet to LambdaRank to LambdaMART: An overview},
  author={Burges, Christopher JC},
  journal={Learning},
  volume={11},
  pages={23--581},
  year={2010}
}

@article{stone1974cross,
  title={Cross-validatory choice and assessment of statistical predictions},
  author={Stone, Mervyn},
  journal={Journal of the royal statistical society: Series B (Methodological)},
  volume={36},
  number={2},
  pages={111--133},
  year={1974},
  publisher={Wiley Online Library}
}

@article{huang2025back,
  title={Back to the Basics: Rethinking Issue-Commit Linking with LLM-Assisted Retrieval},
  author={Huang, Huihui and Widyasari, Ratnadira and Zhang, Ting and Irsan, Ivana Clairine and Shi, Jieke and Ang, Han Wei and Liauw, Frank and Ouh, Eng Lieh and Shar, Lwin Khin and Kang, Hong Jin and others},
  journal={arXiv preprint arXiv:2507.09199},
  year={2025}
}

@inproceedings{reimers2019sentence,
  title={Sentence-bert: Sentence embeddings using siamese bert-networks},
  author={Reimers, Nils and Gurevych, Iryna},
  booktitle={Proceedings of the 2019 conference on empirical methods in natural language processing and the 9th international joint conference on natural language processing (EMNLP-IJCNLP)},
  pages={3982--3992},
  year={2019}
}

@inproceedings{akiba2019optuna,
  title={Optuna: A next-generation hyperparameter optimization framework},
  author={Akiba, Takuya and Sano, Shotaro and Yanase, Toshihiko and Ohta, Takeru and Koyama, Masanori},
  booktitle={Proceedings of the 25th ACM SIGKDD international conference on knowledge discovery \& data mining},
  pages={2623--2631},
  year={2019}
}

@article{ke2017lightgbm,
  title={Lightgbm: A highly efficient gradient boosting decision tree},
  author={Ke, Guolin and Meng, Qi and Finley, Thomas and Wang, Taifeng and Chen, Wei and Ma, Weidong and Ye, Qiwei and Liu, Tie-Yan},
  journal={Advances in neural information processing systems},
  volume={30},
  year={2017}
}

@article{woolson2007wilcoxon,
  title={Wilcoxon signed-rank test},
  author={Woolson, Robert F},
  journal={Wiley encyclopedia of clinical trials},
  pages={1--3},
  year={2007},
  publisher={Wiley Online Library}
}

@article{cliff1993dominance,
  title={Dominance statistics: Ordinal analyses to answer ordinal questions.},
  author={Cliff, Norman},
  journal={Psychological bulletin},
  volume={114},
  number={3},
  pages={494},
  year={1993},
  publisher={American Psychological Association}
}

@article{akhavan2025linkanchor,
  title={LinkAnchor: An Autonomous LLM-Based Agent for Issue-to-Commit Link Recovery},
  author={Akhavan, Arshia and Hosseinpour, Alireza and Heydarnoori, Abbas and Keshani, Mehdi},
  journal={arXiv preprint arXiv:2508.12232},
  year={2025}
}

@inproceedings{brindescu2014centralized,
  title={How do centralized and distributed version control systems impact software changes?},
  author={Brindescu, Caius and Codoban, Mihai and Shmarkatiuk, Sergii and Dig, Danny},
  booktitle={Proceedings of the 36th international conference on Software Engineering},
  pages={322--333},
  year={2014}
}

@misc{abhi_repo,
  author       = {Abhishek Kumar},
  title        = {LinkRank: Learning-to-Rank for One-to-Many Issue--Commit Traceability Recovery},
  year         = {2026},
  howpublished = {\url{https://github.com/merealone2516/LinkRank}},
  note         = {GitHub repository}
}

\end{document}